\documentclass[fleqn,usenatbib]{mnras}

\usepackage{newtxtext,newtxmath}

\usepackage[T1]{fontenc}

\DeclareRobustCommand{\VAN}[3]{#2}
\let\VANthebibliography\thebibliography
\def\thebibliography{\DeclareRobustCommand{\VAN}[3]{##3}\VANthebibliography}

\usepackage{graphicx}	
\usepackage{amsmath}	
\usepackage{array}
\usepackage[table]{xcolor}

\title[On the Ocean Conditions of Hycean Worlds]{On the Ocean Conditions of Hycean Worlds}

\author[F. E. Rigby $\&$ N. Madhusudhan]{
Frances E. Rigby$^{1}$\thanks{E-mail: fer29@ast.cam.ac.uk}
and Nikku Madhusudhan$^{1}\thanks{E-mail: nmadhu@ast.cam.ac.uk}$
\\
$^{1}$Institute of Astronomy, University of Cambridge, Madingley Road, Cambridge CB3 0HA, UK\\
}

\date{Accepted 29th January 2024. Received 26th January 2024; in original form 9th October 2023.}

\pubyear{2024}

\begin{document}
\label{firstpage}
\pagerange{\pageref{firstpage}--\pageref{lastpage}}
\maketitle

\begin{abstract}
Recent studies have suggested the possibility of Hycean worlds, characterised by deep liquid water oceans beneath H$_2$-rich atmospheres. These planets significantly widen the range of planetary properties over which habitable conditions could exist. We conduct internal structure modelling of Hycean worlds to investigate the range of interior compositions, ocean depths and atmospheric mass fractions possible. Our investigation explicitly considers habitable oceans, where the surface conditions are limited to those that can support potential life. The ocean depths depend on the surface gravity and temperature, confirming previous studies, and span 10s to $\sim$1000 km for Hycean conditions, reaching ocean base pressures up to $\sim$6$\times$10$^4$ bar before transitioning to high-pressure ice. We explore in detail test cases of five Hycean candidates, placing constraints on their possible ocean depths and interior compositions based on their bulk properties. We report limits on their atmospheric mass fractions admissible for Hycean conditions, as well as those allowed for other possible interior compositions. For the Hycean conditions considered, across these candidates we find the admissible mass fractions of the H/He envelopes to be $\lesssim$10$^{-3}$. At the other extreme, the maximum H/He mass fractions allowed for these planets can be up to $\sim$4-$8\%$, representing purely rocky interiors with no H$_2$O layer. These results highlight the diverse conditions possible among these planets and demonstrate their potential to host habitable conditions under vastly different circumstances to the Earth. Upcoming JWST observations of candidate Hycean worlds will allow for improved constraints on the nature of their atmospheres and interiors. 
 
\end{abstract}




\section{Introduction}

Earth is the only environment in the universe known to host life. Therefore it is logical that the search for habitable exoplanets and biosignatures began by focusing on Earth-like, rocky exoplanets \citep[e.g.][]{Kasting1993,Meadows2018,Ramirez2018b}. However the rapidly increasing number and diversity of detected exoplanets has prompted wider considerations for habitability studies and candidates for biosignature detections \citep[e.g.][]{Madhusudhan2021}. Another related element is identifying conducive targets for atmospheric characterisation and biosignature detections based on our current and upcoming facilities. The sub-Neptune regime spans planets with radii $\sim$$1-4\ \mathrm{R_\oplus}$, between those of Earth and Neptune. The larger sizes of sub-Neptunes compared to small, rocky, Earth-like exoplanets makes these planets more conducive to atmospheric characterisation via transit spectroscopy. Similar interest arises for planets transiting M dwarfs \citep[e.g.][]{Wunderlich2019,Tremblay2020}. Numerous  sub-Neptunes transiting bright (J < 10 mag) M dwarfs have been discovered by recent transit surveys \citep[e.g.][]{Ricker2015,Gunther2019,Hardegree2020,Cloutier2020}. Spectroscopic observations of these planets with the James Webb Space Telescope (JWST) provide exciting opportunities for detailed atmospheric characterisation and the potential for biosignature searches \citep[e.g.][]{Madhusudhan2023b}. 

A new type of habitable exoplanet within the sub-Neptune regime was recently proposed by \citet{Madhusudhan2021}, known as Hycean worlds. These temperate planets are characterised by their deep H$_2$O oceans and H$_2$-rich atmospheres, and provide a new avenue for habitability studies. The impetus for Hycean worlds came from the study of the habitable-zone sub-Neptune K2-18~b \citep{Madhusudhan2020}. By coupling atmosphere and interior models, they placed constraints on the composition and surface conditions of K2-18~b, finding that this planet could host liquid water at its surface beneath an H$_2$-rich atmosphere. Hycean worlds were shown to broaden the commonly considered limits of habitability, in mass, radius, temperature and orbital distance. Due to their H$_2$ atmospheres, the Hycean habitable zone is significantly wider than the terrestrial habitable zone. Importantly, both the larger size and larger atmospheric scale height of Hyceans relative to rocky planets of similar mass make them more promising targets for atmospheric spectroscopy and potential biosignature detections. Recent analysis of JWST transmission spectra of K2-18~b by \citet{Madhusudhan2023b} revealed strong detections of both CH$_4$ and CO$_2$, with a lack of key molecules including NH$_3$, suggesting the presence of a surface ocean, based on chemical arguments \citep{Yu2021,Hu2021,Tsai2021, Madhusudhan2023a}. We are in the exciting position where detailed atmospheric data for multiple temperate sub-Neptunes orbiting M dwarfs is soon to be accessible, with observations already being carried out in Cycle 1 (for example, K2-18~b in GO programs 2722 and 2372), with more sub-Neptune targets being observed in other programs in Cycles 1 and 2.

Internal structure modelling plays an important role in characterising sub-Neptunes, relating observable properties to possible interior compositions \citep[e.g.][]{Rogers2010b,Valencia2013,Dorn2017,  Madhusudhan2020,Nixon2021,Huang2022}. The sub-Neptune population is thought to contain a diverse range of interior compositions with varying proportions of volatiles including H$_2$O and H/He, with volatile-rich and volatile-poor populations separated by the radius valley \citep[e.g.][]{Fulton2017}. One of the key challenges in internal structure modelling of sub-Neptunes is compositional degeneracy, with a range of compositions usually able to explain a planet's observed mass and radius. In the era of JWST, interior constraints can be drastically improved via information revealed about exoplanet atmospheres, allowing us to gain more insight into the nature of these planets than ever before. Atmospheric observations provide the key in breaking the aforementioned degeneracies between possible interior compositions \citep{Madhusudhan2020}. For example, the presence of a steam atmosphere can be ruled out and an H$_2$-rich atmosphere can be established through transmission spectroscopy \citep[e.g.][]{Benneke2019,Tsiaras2019,Madhusudhan2020,MikalEvans2023}. Even having established the presence of an H$_2$-rich atmosphere, Hycean compositions can be degenerate with rocky worlds with thick H/He envelopes, and with mini-Neptunes with large fractions of ices -- see Figure~1 in \citet{Madhusudhan2023a}. Precise atmospheric data can allow the possible identification of Hycean worlds among the sub-Neptune population. Based on photochemical models, \citet{Madhusudhan2023a} outlined a framework for diagnosing a Hycean world using retrieved chemical abundances, considering the effect of the surface ocean on the atmospheric composition. Notably in recent work, \citet{Madhusudhan2023b} infer the possibility of a liquid H$_2$O surface on Hycean candidate K2-18~b.  

The interiors of Hycean worlds can possess substantial fractions of H$_2$O, up to $90\%$ by mass \citep{Madhusudhan2021}, compared to $\sim$$0.02\%$ for the Earth \citep{Hidenori2016}. Accurate internal structure modelling of H$_2$O-rich planets, including Hycean worlds, requires taking into account the thermal behaviour of H$_2$O across the complex phase diagram \citep[e.g.][]{Thomas2016a,Mousis2020,Huang2021,Nixon2021}. This is achieved via a pressure and temperature dependent equation of state (EOS), which tends to be compiled from several data sources valid for different phases and/or regions of $P$-$T$ space \citep[e.g.][]{Thomas2016a,Nixon2021,Haldemann2020}. At high pressures H$_2$O forms high-pressure ices, which would occur deep in Hycean interiors -- specifically, these are ice VI, VII and ice X \citep{Noack2016,Nixon2021}. The behaviour of high-pressure ices remains uncertain due to a lack of extensive experimental data, which can affect the accuracy of internal structure models for water-rich planets \citep[e.g.][]{Huang2021}. 

Several recent studies have investigated the internal structures and ocean depths of temperate water-rich sub-Neptunes. \citet{Nixon2021} conducted internal structure modelling to examine the H$_2$O phase structure across a wide range of masses, compositions and surface conditions of such planets. Other studies have also explored the range of ocean depths on sub-Neptunes, under more specific assumptions \citep[e.g.][]{Leger2004,Sotin2007,Alibert2014,Noack2016}. For instance, \citet{Alibert2014} investigated cases avoiding high-pressure ice layers, and \citet{Sotin2007} assumed a fixed surface temperature of $300$~K. The results of \citet{Nixon2021} highlight the wide range of parameter space over which liquid water could exist at the surface of a planet, which suggests a diverse range of planets could host Hycean conditions. They also investigated the key factors in determining ocean depth, finding this to be surface gravity and ocean base pressure (and hence surface temperature, due to the adiabatic temperature structure), and constrained the range of ocean depths possible across a wide phase space as functions of these. 

In this study, we focus on Hycean worlds, requiring habitable pressures and temperatures at the interface between the ocean and the H$_2$-rich atmosphere. Based on previous studies, we expect the depths to reach hundreds of times the depth on Earth. For example, \citet{Nixon2021} find that depending on the planet's interior composition and mass, at surface temperatures of $300$~K a planet can host oceans between $30-500$ km deep. Similarly, depths of $\sim$$100-400$~km were found to be possible for the canonical Hycean world based on K2-18~b in \citet{Madhusudhan2023a}, assuming habitable surface conditions. Given the significant H$_2$O mass fractions expected in Hycean worlds the ocean base would occur at the transition to high-pressure ice, as opposed to a rocky ocean floor as on Earth, which could have implications for their habitability \citep{Maruyama2013,Noack2016,Journaux2020b,Madhusudhan2023a}. For example, the thick mantle of high-pressure ice on Hycean worlds would prevent the weathering of the rocky core below, necessitating alternative methods of nutrient enrichment in the oceans. \citet{Madhusudhan2023a} explored the possible chemical conditions on Hycean worlds, identifying feasible pathways to concentrate bioessential elements in Hycean oceans. These include atmospheric condensation, external delivery and convective transport from the rocky core across the ice mantle.

\begin{figure*}
    \centering
    \includegraphics[width=1.9\columnwidth]{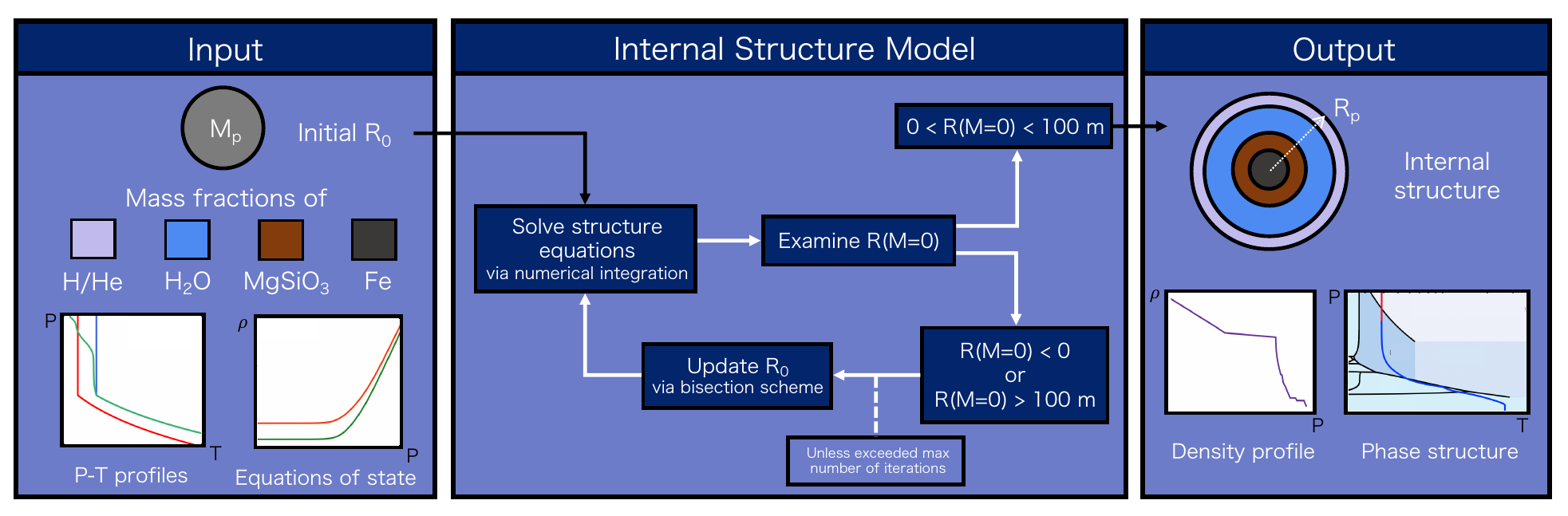}
    \caption{Diagram of the HyRIS internal structure code architecture. The code takes inputs of planetary mass and the mass fractions of its interior layers, and solves the planetary structure equations via numerical integration to output the planet radius. Equations of state and $P$-$T$ profiles are adopted for each layer.}
    \label{fig:codediagram}
\end{figure*}

The lack of sub-Neptunes in our own solar system and the ubiquity of exoplanets in this regime mean that there is a wealth of information to be gained from their study, on planet formation and evolution, in addition to habitability. The radius valley is the observed dearth of planets in the radius range $\sim$$1.5-2\ \mathrm{R_\oplus}$, which separates two subpopulations with bimodal peaks at $\sim$$1.4\ \mathrm{R_\oplus}$ and $\sim$$2.4\ \mathrm{R_\oplus}$ \citep{Fulton2017,Fulton2018,Petigura2020}. The low radius peak is commonly accepted to be largely rocky super-Earths, while the characteristics of the planets in the second peak remain debated, and link to their formation/evolution mechanism. The position of the radius valley is dependent on both orbital period and stellar mass \citep{Fulton2018}, with the trend with stellar insolation reversing for M dwarfs compared to more massive stars \citep{Cloutier2020b}. Formation mechanisms for the radius valley remain debated, and fall broadly into two categories. The first category relies on atmospheric mass loss, such that super-Earths are the envelope-stripped remnants of mini-Neptunes without large water ice fractions -- i.e. largely rocky cores with an H$_2$-rich atmosphere \citep[e.g.][]{Lopez2012,Owen2013,Gupta2019,Rogers2021}. Two methods of mass loss are generally considered, photoevaporation \citep[e.g.][]{Owen2013,Owen2017} and core-powered \citep[e.g.][]{Gupta2019}. The second category places a larger emphasis on inherent differences in composition, with the larger radius peak containing planets with large H$_2$O components \citep[e.g.][]{Zeng2019,Mousis2020,Venturini2020,Izidoro2022}. The nature and formation of the sub-Neptune population remains an open question, with the permitted mass fraction for an H$_2$-rich atmosphere of a sub-Neptune varying by the assumed formation/evolution mechanism. Identifying the envelope mass fractions of the sub-Neptune population is hence important in testing these mechanisms. Furthermore, constraining the composition requirements of Hycean worlds is an important step for investigating the formation/evolution pathway for this class of planet, which has yet to be studied in detail.

In this study we present an analysis of the possible conditions on Hycean worlds using a selection of Hycean candidates due to be observed with JWST, including TOI-270~d, TOI-732~c, TOI-1468~c, K2-18~b and LHS 1140~b. We use our internal structure model to estimate the range of possible ocean depths for these planets as Hycean worlds, and the maximum mass fraction in H/He to allow habitable conditions. We consider another end-member scenario, of a rocky planet with a deep H$_2$-rich atmosphere and no ocean, to constrain the overall upper limit for H/He mass fraction. We discuss the implications of the envelope mass fractions for sub-Neptune formation/evolution, and the effects of Hycean conditions on the observable properties of these planets.

\section{Methods}

In this section we present an internal structure model for sub-Neptunes, HyRIS, with special application to Hycean worlds. We first describe the model and the functionality of the code. The specific assumptions made for our purpose of studying candidate Hycean worlds are outlined. These include the adopted equation of state (EOS) and temperature profile used to describe each planetary component layer. 

\subsection{Internal Structure Model}

Over the past two decades, a number of studies have developed internal structure models to study planetary interiors \citep[e.g.][]{Leger2004,Fortney2007,Seager2007,Sotin2007,Valencia2007,Rogers2010a,Madhusudhan2012,Zeng2013,Thomas2016a,Brugger2017,Madhusudhan2020,Nixon2021,Huang2022}. Here we describe HyRIS, our internal structure model for sub-Neptunes, and how this is customised for the study of Hycean worlds. 

The model solves the planetary structure equations under the assumption of spherical symmetry. The planetary structure equations are the mass continuity equation,

\begin{equation}
    \frac{dR}{dM} = \frac{1}{4\pi R^{2}\rho}
    \label{eqn:masscontinuity}
\end{equation}

\noindent
and the equation for hydrostatic equilibrium,

\begin{equation}
    \frac{dP}{dM} = -\frac{GM}{4\pi R^{4}}
    \label{eqn:hydrostaticeq}
\end{equation}

\noindent
where for a spherical shell, $R$ is the radius, $M$ is the mass enclosed, $\rho$ is the density and $P$ is the pressure. The EOS gives the density as a function of pressure and temperature, $\rho = \rho(P,T)$. The choice of EOS and temperature profile, $T=T(P)$, for each layer are outlined in Sections \ref{EOSs} and \ref{PTprofiles}. In the default set-up of HyRIS, four differentiated layers are considered -- a H/He envelope, a pure H$_2$O layer, a silicate mantle and an iron core, as are commonly considered for sub-Neptunes \citep[e.g.][]{Rogers2010a,Madhusudhan2020,Nixon2021}.

In a similar method to \citet{Nixon2021}, the structure equations are solved using a using a fourth-order Runge-Kutta numerical integration procedure. The boundary conditions are chosen to be at the planet's surface since these are associated with atmospheric observables. The model solves for the planet radius $R_\mathrm{p}$, taking inputs of planet mass $M_{\mathrm{p}}$, mass fractions of its constituents $x_{\mathrm{i}} = M_{\mathrm{i}}/M_{\mathrm{p}}$, and photospheric pressure $P_0$ and temperature $T_0$. If no envelope is included, the boundary $P_0$ and $T_0$ will be at the planet's surface. The integration procedure works inwards from the outside of the planet, stepping through decreasing enclosed mass $M$, where the size of mass step $dM$ is adjusted according to the current $M$ and $\rho$. $R_{\mathrm{p}}$ is obtained via a bisection root-finding method with a convergence condition of a radius value of less than $100$ m at zero enclosed mass, $0 < R(M=0) < 100\ \mathrm{m}$. In addition to $R_{\mathrm{p}}$ and the thickness of each planetary layer, the interior density profile and H$_2$O phase structure can be output if required. The model architecture is summarised in Figure~\ref{fig:codediagram}. 

The EOSs adopted for the different materials are described in Section \ref{EOSs}. However HyRIS is flexible, with the nature of each planetary layer and the associated EOS able to be easily modified. For instance, miscibility of the H$_2$O and H/He layers can be included (e.g. Section \ref{MixedEnvelopes}). See Section \ref{FutureIS} for further discussion of possible developments and adaptations to the model.

\begin{figure*}
    \centering
    \includegraphics[width=0.95\columnwidth]{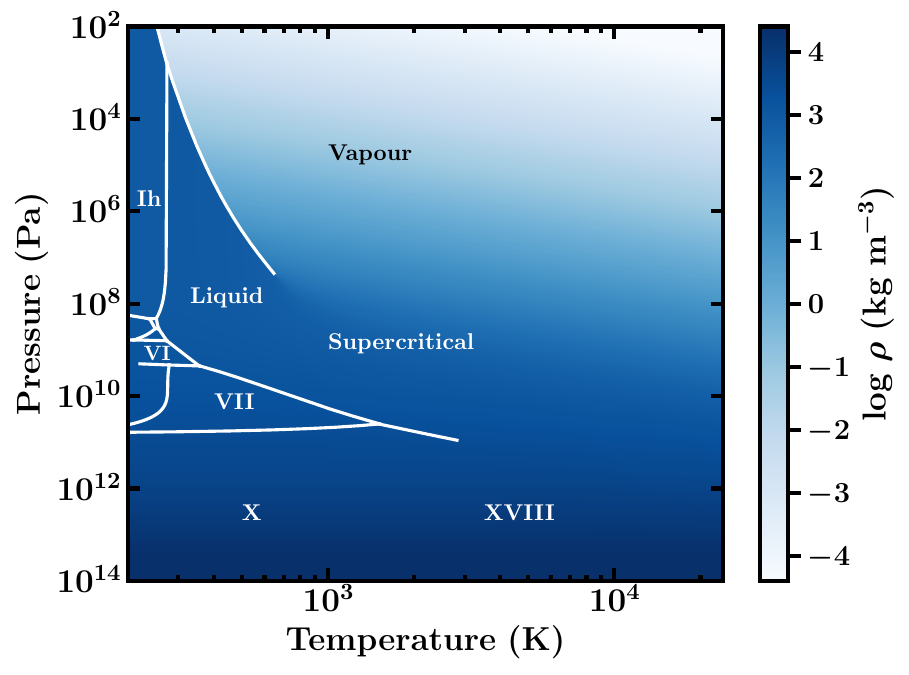}
    \includegraphics[width=\columnwidth]{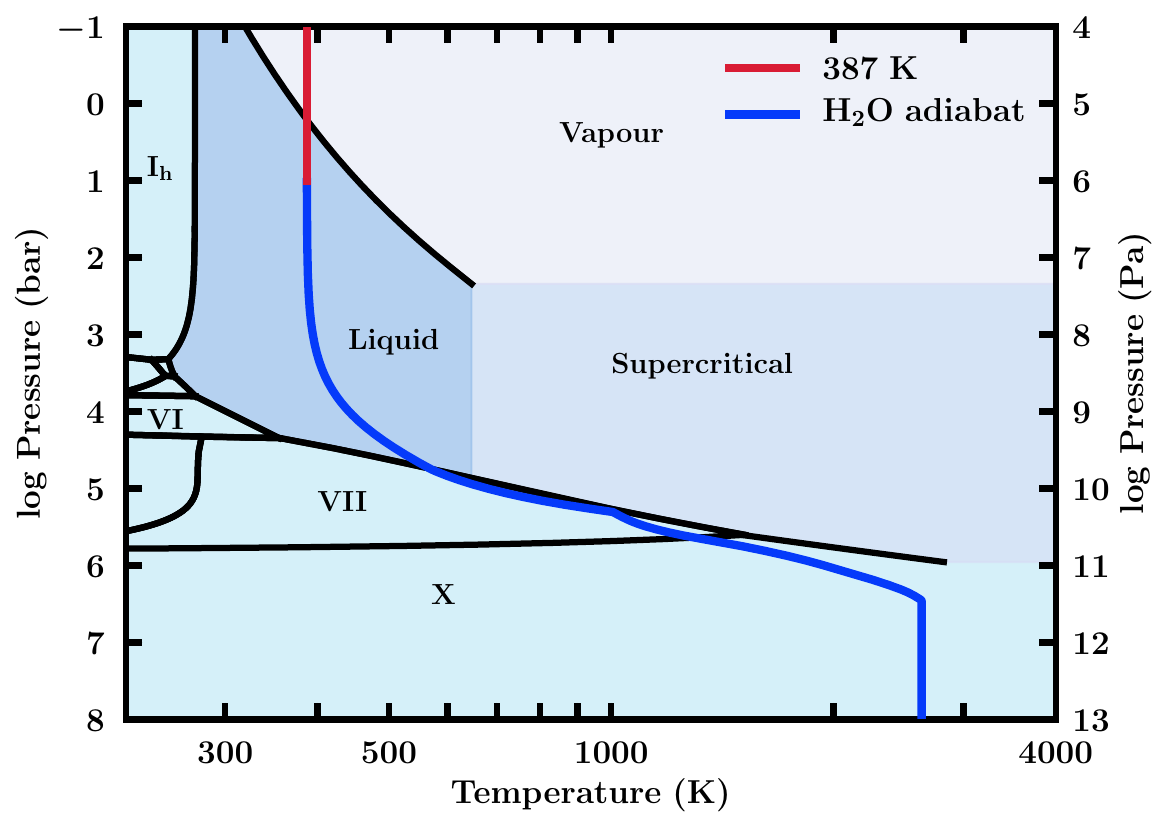}
    \caption{Left: Phase diagram and EOS for H$_2$O used in our model. The phase diagram is constructed from \citet{Dunaeva2010} and \citet{Wagner2002}, and the sources for the EOS are outlined in Table~\ref{table:eossources}. Right: Example H$_2$O adiabat with HHB at $387$~K and $\sim10$ bar. This is the profile for the possible interior of TOI-270~d shown in Figure~\ref{fig:ocean}, with $387$~K the equilibrium temperature with $A_\mathrm{B}=0$  for TOI-270~d. The red line shows the atmospheric profile, which follows an isotherm at these pressures. The black lines and background shading show the phase diagram of H$_2$O.}
    \label{fig:h2oeos}
\end{figure*}

\subsubsection{Exploring Hycean conditions}

For this study, HyRIS has been customised to explore the parameter space for Hycean worlds, and facilitate quick extraction of useful quantities. We automate the extraction of $1\sigma$ solutions of internal structures for the $M_\mathrm{p}$ and $R_\mathrm{p}$ measurements for a given planet from a large number of interior model executions across the full parameter space of possible mass fractions. We further automate the determination of Hycean solutions from the $1\sigma$ solutions. As discussed above, the model can output the density profile and H$_2$O phase structure, along with the $R_\mathrm{p}$ and thickness of each planetary layer. For a given solution, if the surface is found to lie in the liquid phase of H$_2$O, the ocean depth is calculated. For Hyceans, with liquid surfaces with temperatures up to $\sim$$400$ K, the base of the ocean will be high-pressure ice. Hotter surfaces can lead to supercritical oceans, as shown in \citet{Nixon2021}.

\subsection{Equations of State} \label{EOSs}

For each planetary layer considered which are typically H/He, H$_2$O, silicates and iron, we require an EOS to describe the pressure and temperature dependent density variation within each layer. We describe the choice of EOS for each layer and, where relevant, the process to compile them. 

\begin{figure*}
    \centering
    \includegraphics[width=\columnwidth]{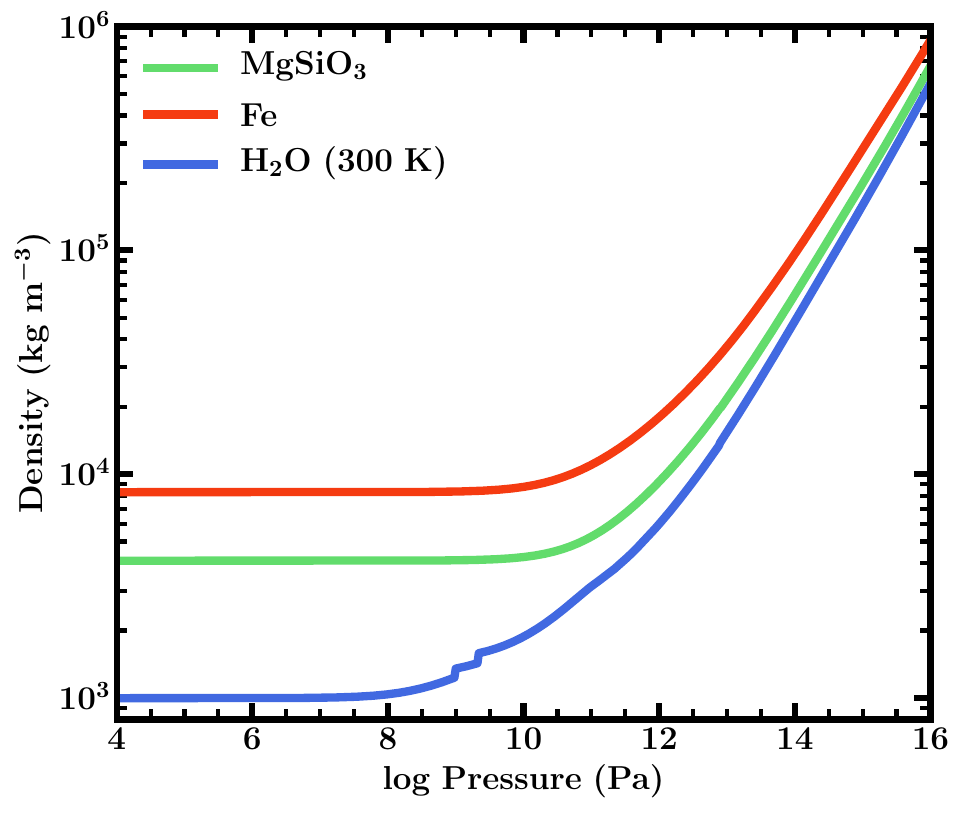}
    \includegraphics[width=\columnwidth]{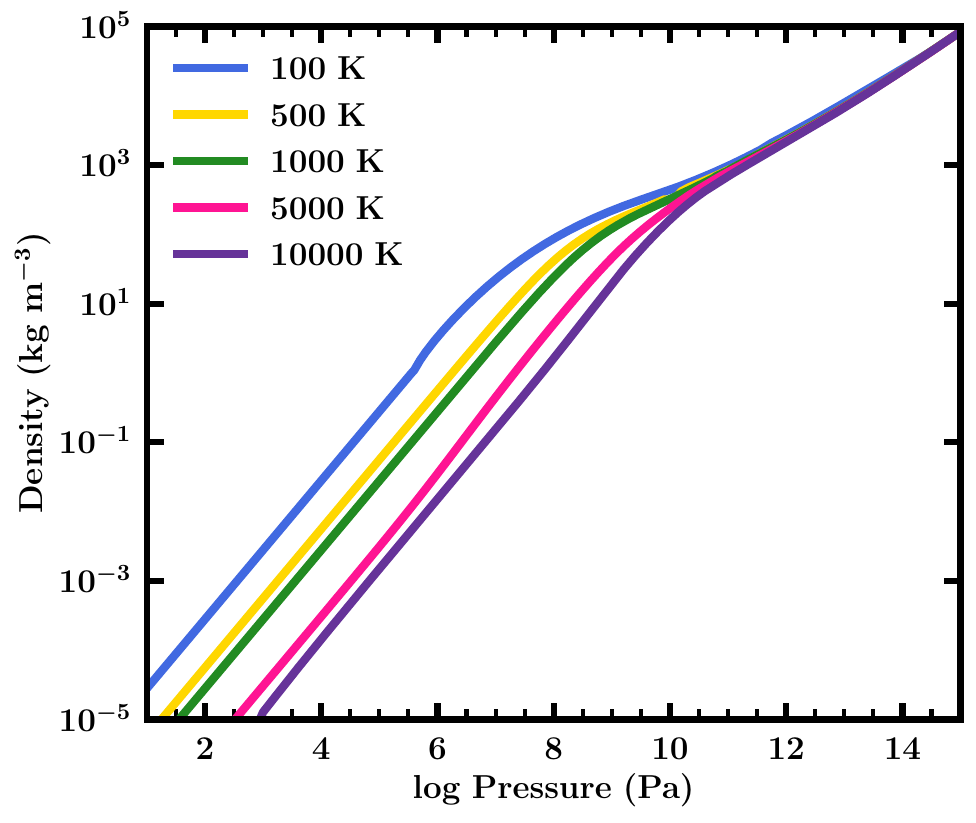}
    \caption{Left: The temperature-independent EOS for each of MgSiO$_3$ and Fe used in our model \citep[from][]{Seager2007}. The equivalent EOS for H$_2$O from \citet{Seager2007} is also shown for comparison. Right: The EOS for H/He used in our model \citep[from][]{Chabrier2019}, shown for different isotherms.}
    \label{fig:eoss}
\end{figure*}

\subsubsection{H$_2$O} \label{H2OEOS}

\begin{table}
\begin{center}
\caption{Sources and regions of validity for the components of our H$_2$O EOS, shown in Figure~\ref{fig:h2oeos}.} 
\setlength{\arrayrulewidth}{1.0pt}
\begin{tabular}{ wc{3.8cm}  wc{4.2cm} }

 \hline
 \rowcolor{lightgray} \textbf{EOS and Source} & \textbf{Validity}\\ 
 \hline{}
 IAPWS-1995, & Vapour, liquid, supercritical\\ 
 \citep{Wagner2002} & \\

 \citet{French2009} & Supercritical, ice VII, ice X, ice XVIII, \\
 & plasma. $1000-24000$~K, \\
 & $1.86\times10^{9}-9.87\times10^{12}$ Pa \\

 \citet{Feistel2006} & Ice Ih.\\
 & $0-273$~K, $0-10^{8}$ Pa\\ 

 \citet{Journaux2020a} & Ices II, III, V, VI\\ 

 \citet{Fei1993} & Ice VII\\

 \citet{Fei1993}, \citet{Klotz2017} & Ice VIII \\

 Thomas-Fermi-Dirac (TFD), & $P > 7.686\times10^{12}$ Pa\\
 \citep{Salpeter1967} & \\

 \citet{Seager2007} & Remaining high-pressure regions\\

 IAPWS-1995 extrapolation, & Remaining regions, (low-pressure\\ 
 \citep{Wagner2002} &  and high-temperature vapour)\\
 \hline
\end{tabular}
\label{table:eossources}
\end{center}
\end{table}

We use a temperature-dependent EOS for pure H$_2$O, compiled following a similar approach to \citet{Thomas2016a} and \citet{Nixon2021}. Our EOS is valid for temperatures in the range $200 - 24000$~K and pressures $10^2 - 10^{22}$ Pa ($10^{-3}-10^{17}$ bar). It is comprised of a number of different sources valid for certain phases and/or regions of $P$-$T$ space. These sources will be described below, and are summarised in Table~\ref{table:eossources}. We use phase-boundaries from \citet{Dunaeva2010}, in addition to the liquid-vapour boundary from \citet{Wagner2002}. 

For the liquid and vapour phases and some parts of the supercritical phase, we use the EOS from the International Association for the Properties of Water and Steam (IAPWS) \citep{Wagner2002}, referred to as the IAPWS-1995 formulation. This EOS has been well tested experimentally. We use the functional form of the IAPWS-1995 formulation, which is calculated directly from the Helmholtz free energy, to give pressure as a function of density and temperature, $P=P(\rho,T)$. This requires a bounded root-finding procedure to be carried out to obtain $\rho=\rho(P,T)$ for each phase. 

We use the EOS of \citet{French2009}, the data for which covers multiple phases, including supercritical, high-pressure ices (ice VII, X and XVIII) and plasma. This EOS is based on quantum molecular dynamics simulations, and has since been experimentally validated by \citet{Knudson2012}. The EOS for remaining regions of the supercritical phase is an extrapolation of the IAPWS-1995 EOS.

For the majority of ice VII (the high-temperature region is covered by the \citet{French2009} EOS), we use a functional EOS in the form of a Vinet EOS with a thermal correction from \citet{Fei1993}. The Vinet EOS is given by

\begin{equation}
    P = 3B_{0}\eta^{\frac{2}{3}}\left(1-\eta^{-\frac{1}{3}}\right)\exp{\left[\frac{3}{2}\left(B^{'}_{0}-1\right)\left(1-\eta^{-\frac{1}{3}}\right)\right]}
    \label{eqn:Vinet}
\end{equation}

where $\eta=\rho/\rho_{0}$, $\rho_{0}$ is the ambient density, $B_0=\rho\left(\partial P/\partial \rho\right)|_{T}$ is the isothermal bulk modulus and $B^{'}_{0}$ its pressure derivative. The thermal correction is given 

\begin{equation}
    \rho\left(P,T\right) = \rho_{0}\left(P,T_{0}\right) \left[\exp\left(\int_{T_{0}}^{T}\alpha\left(P,T\right)dT\right)\right]^{-1}
    \label{eqn:Vinetthermalcorrection}
\end{equation}

\noindent
where $T_0$ is the ambient temperature, here $300$ K \citep{Fei1993}. The thermal expansion coefficient $\alpha$ here is 

\begin{equation}
    \alpha\left(P,T\right) = \alpha_{0}\left(T\right)\left[1+\frac{B^{'}_{0}}{B_{0}}P\right]^{-\eta}
    \label{eqn:alphaicevii}
\end{equation}

\noindent
where $\alpha_{0}\left(T\right)$ is a linear function of $T$, and $\eta$ is a constant \citep{Fei1993}. The coefficients were experimentally determined via x-ray diffraction \citep{Fei1993}. Via an alternative form of $\alpha_0$ from \citet{Klotz2017}, we also extrapolate this EOS to cover ice VIII. 

For ice Ih we use the functional form of the EOS from \citet{Feistel2006}. Ices II, III, V and VI are also covered by experimental data, from \citet{Journaux2020b} via the SeaFreeze package. 

At high pressures, above $7.686\times10^{7}$ bar ($7.686\times10^{12}$) Pa we adopt a modified Thomas-Fermi-Dirac (TFD) EOS as in \citet{Salpeter1967}. This EOS is temperature independent, due to the minimal effect of temperature in this regime. There remains an intermediate region in pressure space between ice VII and X not covered by experimental data or the TFD EOS. In this region we use the H$_2$O EOS from \citet{Seager2007}. This is comprised of three regimes -- at lower pressures this is a Birch-Murnaghan (BM) EOS \citep{Birch1952} with coefficients from \citet{Hemley1987}, transitioning to density functional theory results with increasing pressure, and finally to a TFD at high pressures. 

Our full H$_2$O EOS and phase diagram are shown in Figure~\ref{fig:h2oeos}. The EOS was constructed based on the regions of validity described above and in Table~\ref{table:eossources}. This is either via the phase boundaries, or by the bounds of the data. 

We validate our H$_2$O EOS against the EOS of \citet{Nixon2021}. We expect these to be almost identical due to the sources used. In Figure~\ref{fig:validation} we show this to be the case, showing the density as a function of pressure for different isotherms. 

For $c_\mathrm{p}$, required for the adiabatic gradient (see Section \ref{PTprofiles}), we use the same sources as the EOS where available \citep[for][]{Wagner2002,Feistel2006,Journaux2020a}. For regions where there is no data for $c_\mathrm{p}$, we adopt the $c_\mathrm{p}$ value of the nearest available point in $P$-$T$ space. $\alpha$ is determined directly from the EOS, as in Equation~\ref{eqn:alpha}.

\subsubsection{Silicates and iron} \label{CoreEOSs}

\begin{table}
\begin{center}
\caption{Values used in the EOSs for the iron and silicate layers, from \citet{Anderson2001} and \citet{Karki2000} respectively.} 
\setlength{\arrayrulewidth}{1.0pt}
\begin{tabular}{ wc{1.0cm}  wc{1.0cm}  wc{1.0cm}  wc{1.2cm}  wc{1.2cm} }
 \hline
 \rowcolor{lightgray} \textbf{Layer} & \textbf{B}$_0$ (\textbf{GPa}) & \textbf{B}$_0^{\prime}$ & \textbf{B}$_0^{\prime \prime}$ (\textbf{GPa}$^{-1}$) & $\mathbf{\rho}$ (\textbf{kg m}$^{-3}$) \\
 \hline
 Fe & 156.2 & 6.08 & N/A & 8300\\ 
 MgSiO$_3$ & 247 & 3.97 & -0.016 & 4100\\ 
 \hline
\end{tabular}
\label{table:coreeosparams}
\end{center}
\end{table}

Since thermal effects within the core and mantle have been found to have a minimal effect on the planetary $M$-$R$ relation \citep[e.g.][]{Grasset2009}, we adopt an isothermal EOS in the silicate mantle and iron core, as is frequently assumed in other internal structure models \citep[e.g.][]{Rogers2011,Thomas2016a}. For this we use the EOS from \citet{Seager2007}. These EOSs are shown in Figure~\ref{fig:eoss}.

The iron core is described by a Vinet EOS \citep{Vinet1989,Anderson2001} for hexagonal close-packed Fe, before similarly transitioning to a TFD EOS \citep{Salpeter1967} at higher pressures.  

For the silicate layer, assumed to be the perovskite phase of MgSiO$_3$, the EOS is in the form of a fourth-order BM EOS \citep{Birch1952,Karki2000}, and a TFD EOS \citep{Salpeter1967} at high pressure. The BM EOS is described by

\begin{equation}
    \begin{split}
    P = \frac{3}{2}B_{0}\left( \eta^{\frac{7}{3}}-\eta^{\frac{5}{3}} \right) \Biggl[ 1+\frac{3}{4}\left( B^{'}_{0}-4 \right)\left( \eta^{\frac{2}{3}}-1 \right) + \\
    \frac{3}{8}B_{0}\left( \eta^{\frac{2}{3}}-1 \right)^{2} \left[ B_{0}B_{0}^{''}+B_{0}^{'}\left( B_{0}^{'}-7 \right)+\frac{143}{9} \right] \Biggr]
    \end{split}
\label{eqn:BME4}
\end{equation}

where $B^{''}_{0}$ is the second pressure derivative of the isothermal bulk modulus. The coefficients for this are from \citet{Karki2000}, and are given in Table~\ref{table:coreeosparams}.

\subsubsection{Hydrogen/Helium} \label{HHeEOS}

We assume a solar helium fraction for the H/He envelope ($Y=0.275$). The EOS from \citet{Chabrier2019} is used, which is valid for $1-10^{22}$ Pa and $100-10^{8}$ K. This EOS is comprised of models applicable at different density and temperature regimes, and the hydrogen and helium EOSs are combined via an additive-volume law. In our temperature range, the models used are from \citet{Saumon1995,Caillabet2011,Chabrier1998} -- see \citet{Chabrier2019} for a full description of the model. This EOS is shown in Figure~\ref{fig:eoss}.

\subsection{Temperature Profiles} \label{PTprofiles}

\begin{figure}
    \centering
    \includegraphics[width=0.9\columnwidth]{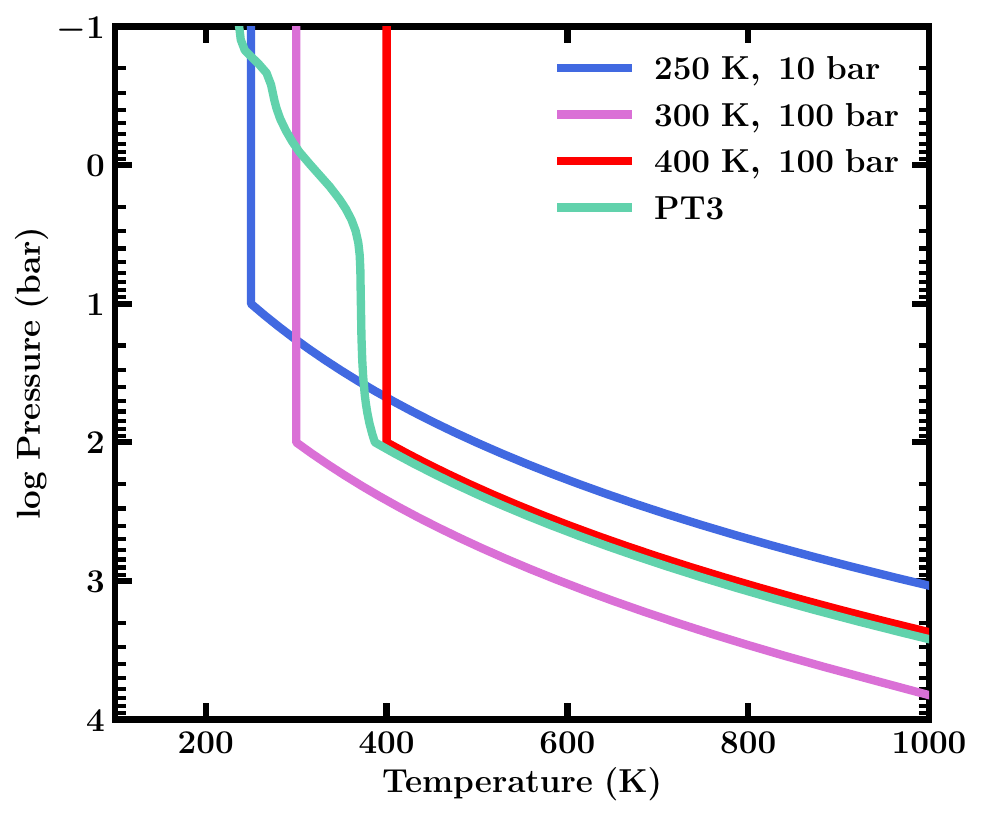}
    \caption{Forms of $P$-$T$ profile used for the H/He envelope. Example isothermal/adiabatic profiles are shown at three different temperatures with $P_\mathrm{rc}$ between $10-100$~bar. PT3 is from \citet{Madhusudhan2023a}, generated via self-consistent modelling with GENESIS \citep{Gandhi2017,Piette2020} for K2-18~b.}
    \label{fig:envelopePT}
\end{figure}

The model considers pressure and temperature dependent EOSs. HyRIS can accommodate user-specified $P$-$T$ profiles for the atmosphere, while in the interior, an adiabatic profile is typically assumed. \citet{Madhusudhan2021} showed that the $P$-$T$ structure expected in the observable H$_2$-rich atmospheres of Hycean worlds around M dwarfs are well approximated by an isothermal profile. We note that the atmospheric compositions of Hycean worlds could be rich in CH$_4$ and CO$_2$ \citep[e.g.][]{Madhusudhan2023b}, which could further affect the temperature structure. \citet{Nixon2021} also showed that $P$-$T$ profiles for sub-Neptunes generated via self-consistent atmospheric modelling with GENESIS \citep{Gandhi2017,Piette2020} could be reasonably approximated by isothermal/adiabatic profiles, with the radiative-convective boundary, $P_\mathrm{rc}$, lying at $\sim$$1-1000$ bar. In the H/He atmosphere, we hence assume an isotherm down to $P_{\mathrm{rc}}=100$ bar, with the $P$-$T$ profile then following an adiabat in the deeper atmosphere. Example profiles are shown in Figure~\ref{fig:envelopePT}. 

Under the assumption of vigorous convection, we similarly assume an adiabatic temperature profile in the H$_2$O layer, as is commonly assumed in internal structure models \citep[e.g.][]{Sotin2007,Nixon2021,Leleu2021}. An example interior adiabat is shown in Figure~\ref{fig:h2oeos}. The adiabatic profiles are described by the adiabatic temperature gradient,

\begin{equation}
    \nabla_{\mathrm{ad}} = \left. \frac{\partial T}{\partial P}\right|_{S} = \frac{\alpha T}{\rho c_p}
    \label{eqn:adiabat}
\end{equation}

\noindent
where $\alpha$ is the coefficient of volume expansion and $c_p$ is the isobaric specific heat capacity. $\alpha$ is derived from the EOS,

\begin{equation}
    \alpha = \frac{1}{V} \left. \frac{\partial V}{\partial T}\right|_{P} = - \left. \frac{\partial \ln{\rho}}{\partial T}\right|_{P}
    \label{eqn:alpha}
\end{equation}

\noindent
where $V$ is the specific volume. 

Alternative temperature profiles can be easily included within HyRIS. For instance, atmospheric temperature profiles generated via self-consistent modelling \citep[e.g.][]{Piette2020} can be used in the place of the isothermal/adiabatic profiles. An example of this is carried out for K2-18~b in Section \ref{K218b}.

\subsection{Model Validation}

\begin{figure*}
    \centering
    \includegraphics[width=0.98\columnwidth]{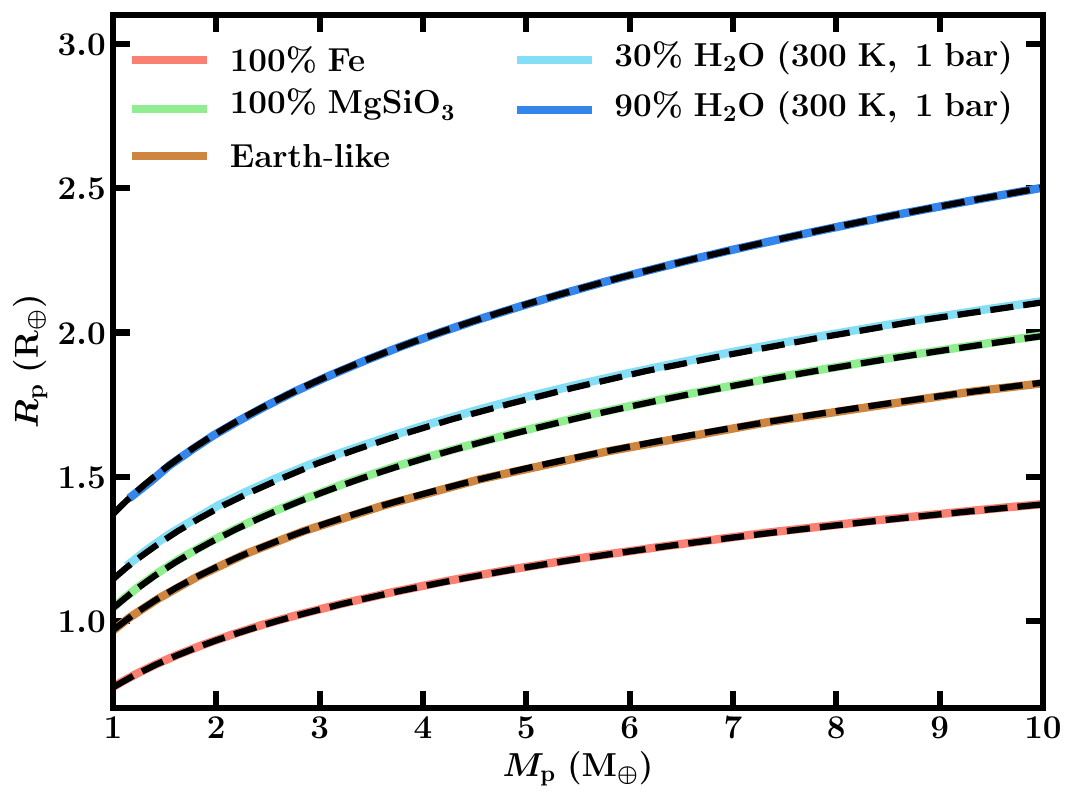}
    \includegraphics[width=\columnwidth]{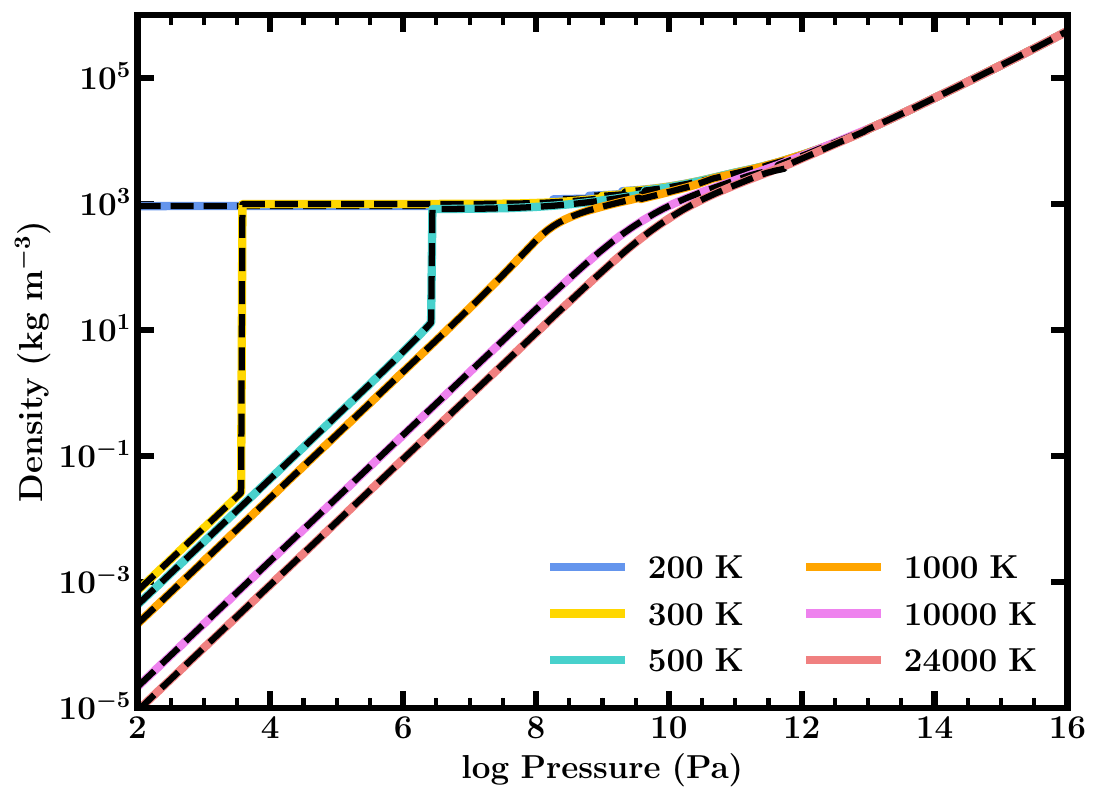}
    \caption{Left: Mass-radius ($M$-$R$) relations produced using our model compared to the results of \citet{Seager2007} and \citet{Nixon2021}. The $100\%$ Fe, $100\%$ MgSiO$_3$ and Earth-like ($2/3$ MgSiO$_3$, $1/3$ Fe) curves are from \citet{Seager2007}. The curves with $30\%$ ($90\%$) H$_2$O and $70\%$ ($10\%$) Earth-like core are from \citet{Nixon2021}, assuming liquid surfaces at $300$ K and $1$ bar. The $M$-$R$ curves produced using our model are shown in black. Right: Comparison of our H$_2$O EOS to the EOS of \citet{Nixon2021}, for several isotherms. Our results are shown in black. These results are almost identical due to the identical sources used for the EOS and phase boundary data.}
    \label{fig:validation}
\end{figure*}

We validate our model against the results of previous studies of sub-Neptune internal structures. In Figure~\ref{fig:validation} we reproduce $M$-$R$ relations from \citet{Seager2007} and \citet{Nixon2021}. The results of our model are shown in black, with the $100\%$ Fe, $100\%$ silicate and Earth-like curves from \citet{Seager2007}, and the $M$-$R$ curves including H$_2$O from \citet{Nixon2021}. The Earth-like composition consists of 2/3 silicate and 1/3 iron. The \citet{Seager2007} cases are taken to be isothermal planets, as the EOSs used are temperature-independent. We expect our results to match \citet{Seager2007} due to the use of identical EOSs, which we see. We also see close agreement with the $M$-$R$ curves from \citet{Nixon2021}, which represent $30\%$ and $90\%$ H$_2$O with an Earth-like core. Henceforth, we use ``core'' to refer to the rocky component of the interior, including both the silicate and iron layers. These cases of $300$ K and $1$ bar surface were chosen as they are representative of a Hycean interior, with the interior $P$-$T$ profile following an adiabat as described in Section \ref{PTprofiles}. In Section \ref{Depths} we compare our model ocean depth results to those of previous studies.

\section{Results}\label{Results}

In this section we present our results for the possible oceanic and interior conditions of Hycean worlds. Firstly we validate our model ocean depths against the results of previous studies \citep{Leger2004,Noack2016,Nixon2021}, before constraining the theoretical range of ocean depths possible on Hycean worlds. We then explore in detail five candidate Hyceans, placing constraints on their possible ocean depths, interior compositions and envelope mass fractions. The envelope mass fractions are determined for the case of a Hycean world and for the case of a rocky world with a thick H/He envelope, the latter representing the overall maximum envelope fraction for these sub-Neptunes.

\subsection{Ocean Depths on Hycean Worlds}\label{Depths}

Hycean worlds are defined to have $10-90\%$ H$_2$O by mass \citep{Madhusudhan2021}, therefore in all calculations we impose these limits on the H$_2$O mass fraction. To allow for the surface ocean, we require the H/He-H$_2$O boundary (HHB) to lie at pressures and temperatures that support the liquid phase of H$_2$O. The HHB is what we will refer to as the ``surface''. We further place the constraint of habitable surface conditions, requiring surface temperatures $T_\mathrm{HHB}$ of $273-400$~K and pressures $P_\mathrm{HHB}$ of $1-1000$ bar, motivated by the range of conditions supporting life on Earth \citep{Rothschild2001,Merino2019}. We consider core compositions between Earth-like ($33\%$ Fe) and pure Fe. The internal structure model is then applied across the full Hycean phase space, for planets with $1-10\ \mathrm{M_\oplus}$, $10\% \leq x_\mathrm{H_2O} \leq 90\%$, $273 \leq T_\mathrm{HHB} \leq400$ K. For each combination we extract $R_\mathrm{p}$ and the ocean depth.

\begin{figure*}
    \centering
    \includegraphics[width=0.98\columnwidth]{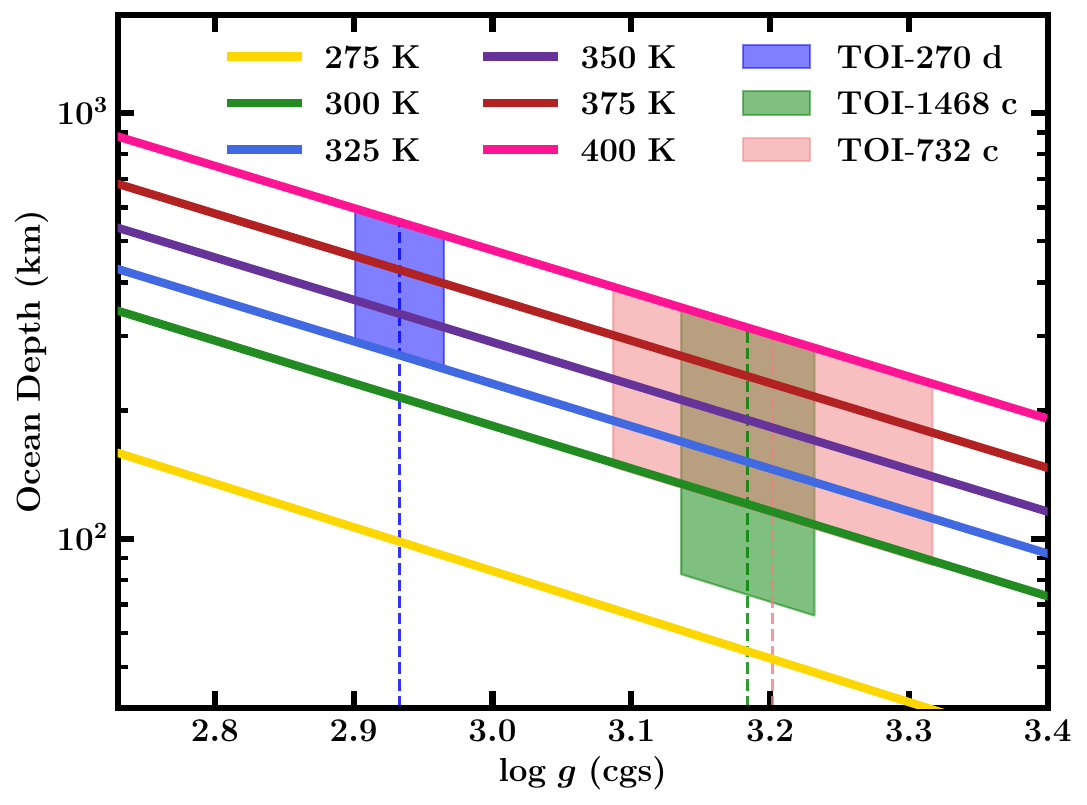}
    \includegraphics[width=0.98\columnwidth]{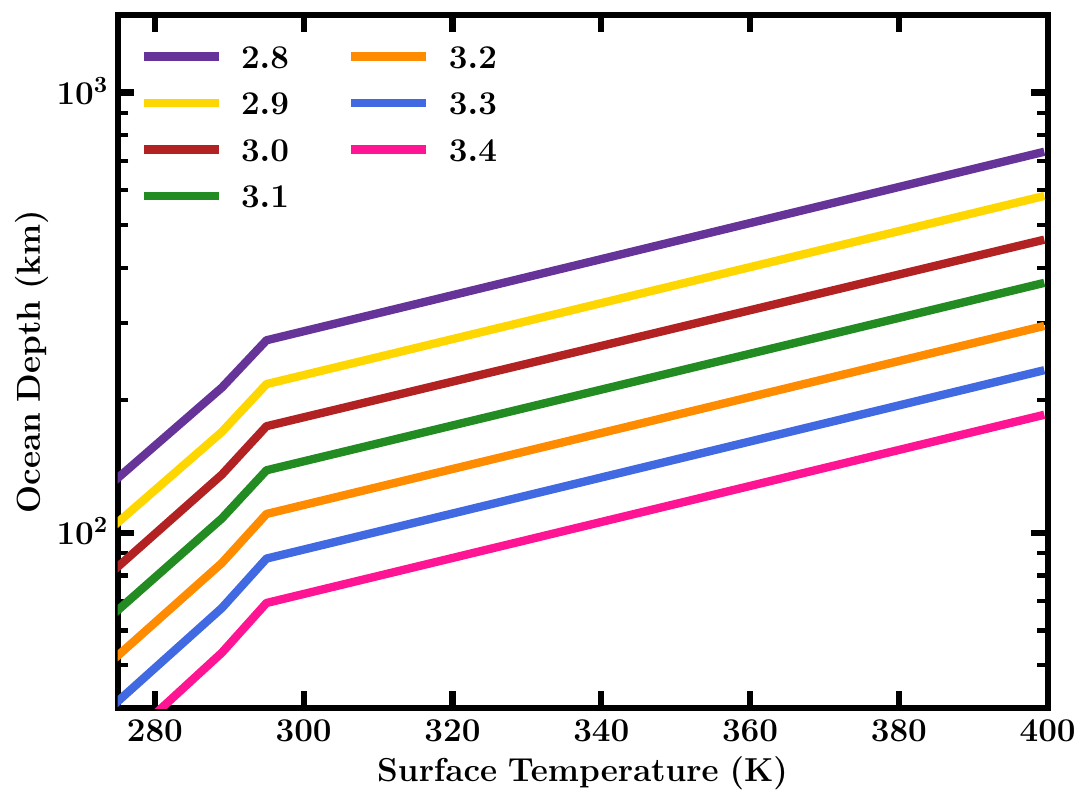}
    \caption{Left (a): Ocean depth against surface gravity for a range of habitable surface temperatures. The shaded regions indicate the estimated range of ocean depths for three Hycean candidates, based on uncertainty in their surface gravity and temperature. The dashed line indicates the median gravity. The mass and radius values adopted for each target are their first entry in Table~\ref{table:targets}. Right (b): Ocean depth against surface temperature for a range of surface gravity values.}
    \label{fig:depths}
\end{figure*}

We first reproduce some previous results in the literature. \citet{Noack2016} found that the maximum ocean depth for a planet varies with the mass, composition and surface temperature of the planet. Similarly, \citet{Nixon2021} showed that the ocean depth is affected by the surface gravity and ocean base pressure. The pressure at the ocean base, where the transition from liquid to high-pressure ice occurs, is fixed by the interior adiabat and hence the surface temperature. In Figure~\ref{fig:depths}a we reproduce these results using our model, showing the inverse proportionality of ocean depth and surface gravity. The surface pressure is fixed at $100$ bar as in \citet{Nixon2021} -- varying this within a reasonable range ($\sim$$1-1000$ bar) does not significantly affect the results, in agreement with \citet{Nixon2021}. In Figure~\ref{fig:depths}b we also show the ocean depth against the HHB temperature for a range of surface gravity values. The change in slope at $\sim$$295$~K corresponds to the transition between an ocean base of ice VI vs ice VII. \citet{Nixon2021} showed that peak depths occur for $T_\mathrm{HHB}=413$~K with the trend reversing for $T_\mathrm{HHB}>413$~K (up to the critical temperature). Therefore, as seen in Figure~\ref{fig:depths}, the trend with temperature is straightforward for our considered temperature range of $273-400$~K, with an increase in $T_\mathrm{HHB}$ causing an increase in ocean depth. We reiterate that in Figure~\ref{fig:depths}, this is the gravity and temperature at the ocean surface ($g_\mathrm{HHB}$, $T_\mathrm{HHB}$), not with the envelope included ($g_0$, $T_0$). As an illustration, for a planet with $M_\mathrm{p}=6\ \mathrm{M_\oplus}$, $50\%$ H$_2$O, $50\%$ Earth-like core and $T_\mathrm{HHB}=303$~K, we obtain a depth of $127$ km. This is in good agreement with the $125$ km found by \citet{Nixon2021} and $133$ km found by \citet{Leger2004}. 

\definecolor{lightred}{RGB}{255, 128, 136}
\definecolor{lightblue}{RGB}{126, 192, 237} 
\definecolor{lightpurple}{RGB}{227, 207, 250} 
\definecolor{lightgreen}{RGB}{102, 209, 142}
\definecolor{lightpink}{RGB}{245, 169, 225}
\definecolor{lightyellow}{RGB}{242, 242, 191}
\setlength{\arrayrulewidth}{1.1pt}
\begin{table*}
\begin{tabular}{  wc{1.4cm}   wc{1.4cm}   wc{1.5cm}   wc{1.0cm}   wc{1.0cm}   wc{0.9cm}   wc{0.9cm}   wc{0.9cm}   wc{0.9cm}   wc{0.7cm}   wc{0.7cm}   wc{0.7cm}  }

     \hline
     \rowcolor{lightgray} \textbf{Planet} & $\mathbf{M_\mathbf{p}/\mathrm{\mathbf{M}}_{\oplus}}$ & $\mathbf{R_{p}/\mathrm{\mathbf{R}}_{\oplus}}$ & \textbf{$\mathbf{T_{eq,0}}$ /K} & \textbf{$\mathbf{T_{eq,0.5}}$ /K} & \textbf{$\mathbf{a}$/AU} & $\mathbf{M_{\star}/\mathrm{\mathbf{M}}_{\odot}}$ & $\mathbf{R_{\star}/\mathrm{\mathbf{R}}_{\odot}}$ & \textbf{$\mathbf{T_{\mathrm{eff}}}$/K} & \textbf{V mag} & \textbf{J mag} & \textbf{Refs}\\
     \hline
     \cellcolor{lightpink} \textbf{K2-18 b} & $8.63\pm1.35$ & $2.610\pm0.087$ & 297 & 250 & 0.153 & 0.45 & 0.44 & 3590 & 13.48 & 9.76 & 1,2\\
     \cellcolor{lightpink}  &  & $2.51_{-0.18}^{+0.13}$ &  &  &  &  & &  &  &  & 3\\
     \cellcolor{lightred} \textbf{TOI-732 c} & $8.60^{+1.60}_{-1.30}$ & $2.30^{+0.16}_{-0.15}$ & 353 & 297 & 0.07673 & 0.40 & 0.37 & 3331 & 13.14 & 9.01 & 4\\
     \cellcolor{lightred} & $6.29^{+0.63}_{-0.61}$ & $2.42\pm0.10$ &  &  &  &  &  &  &  &  & 5\\
     \cellcolor{lightred} & $8.04^{+0.50}_{-0.48}$ & $2.39^{+0.10}_{-0.11}$ &  &  &  &  &  &  &  &  & 6\\    

     \cellcolor{lightgreen} \textbf{TOI-1468 c} & 6.64$^{+0.67}_{-0.68}$ & 2.064$\pm0.044$ & 338 & 284 & 0.0859 & 0.34 & 0.34 & 3496 & 12.50 & 9.34 & 7\\ 

     \cellcolor{lightblue} \textbf{TOI-270 d} & 4.78$\pm0.43$ & 2.133$\pm0.058$ & 387 & 326 & 0.0733 & 0.39 & 0.38 & 3506 & 12.60 & 9.10 & 8,9\\ 
     \cellcolor{lightblue}  & 4.20$\pm0.16$ & 2.19$\pm0.07$ &   &   &   &   &   &   &   &   & 10,11\\ 

     \cellcolor{lightyellow} \textbf{LHS 1140 b} & $6.98\pm0.89$ & $1.727\pm0.032$ & $235$ & $198$ & $0.0936$ & $0.18$ & $ 0.21$ & $3216$ & 14.15 & 9.61 & 12\\     
     \cellcolor{lightyellow} & $6.38^{+0.46}_{-0.44}$ & $1.635\pm0.046 $ &  &  &  & &  &  &  &  & 13\\  
     \cellcolor{lightyellow} & $5.60\pm0.19$ & $1.730\pm0.025 $ &  226 &  &  & &  &  &  &  & 14\\  
     \hline     
\end{tabular}
\caption{Properties of the Hycean candidates considered in this work. Equilibrium temperature values are calculated with $A_\mathrm{B}=0$ and $A_\mathrm{B}=0.5$, and assuming uniform day-night redistribution. References: 1: \citet{Cloutier2019}, 2: \citet{Benneke2019}, 3: \citet{Hardegree2020}, 4: \citet{Cloutier2020}, 5: \citet{Nowak2020}, 6: \citet{Bonfanti2024}, 7: \citet{Chaturvedi2022}, 8: \citet{Gunther2019}, 9: \citet{VanEylen2021}, 10: \citet{MikalEvans2023}, 11: \citet{Kaye2022}, 12: \citet{Ment2019}, 13: \citet{Lillo2020}, 14: \citet{Cadieux2024}.}

\label{table:targets}
\end{table*}

We find that ocean depths on Hycean worlds can range between $10$s of km to $\sim$$1000$ km, depending on the planet's mass, composition and surface conditions. For comparison, the average depth of Earth's ocean is $3.7$ km \citep{Charette2010}, with the deepest part (the Mariana trench) extending to $\sim$$11$ km \citep{Gardner2014}. For the theoretical maximum depth of a Hycean ocean, we find $\sim$$1000$ km -- this end-member case is a $1\ \mathrm{M_\oplus}$ planet with $90\%$ H$_2$O and $T_\mathrm{HHB}=400$~K, i.e. maximal $T_\mathrm{HHB}$ and minimal $\mathrm{log}(g)$. Conversely, the minimum depth is $\sim$$20$ km, for a $10\ \mathrm{M_\oplus}$ planet with $10\%$ H$_2$O, $90\%$ Fe and $T_\mathrm{HHB}=273$~K. 

The calculated ocean depths as a function of surface gravity $g$ and $T_\mathrm{HHB}$, as shown in Figure~\ref{fig:depths}, can be used to estimate the range of ocean depths possible for a given planet. For instance, in Figure~\ref{fig:depths}a we show estimates of the ocean depths possible for TOI-270~d, TOI-1468~c and TOI-732~c, shown by the shaded regions. The $T_\mathrm{HHB}$ range considered is from $T_{\mathrm{eq}}$ at $A_{\mathrm{B}}=0.5$, calculated via Equation \ref{eqn:equilibriumtemp}, up to the maximum of $400$~K. Alternatively, using Figure \ref{fig:depths}b the range of ocean depths can be estimated given a value for the surface gravity. For example, for a surface gravity similar to K2-18~b of $\mathrm{log}\ g=3.1$, the range of ocean depths would be $\sim$$65-380$~km, for HHB temperatures $273-400$~K. In this method, $g$ is implicitly being assumed as constant throughout the atmosphere. In reality, the gravity for each planet would be higher at the ocean surface than at the photosphere, where $R_\mathrm{p}$ is measured from. These represent initial estimates, as not all the solutions in these regions will be permissible for a planet when the gravity in the atmosphere is allowed to vary, and an atmospheric $P$-$T$ profile adopted; these are considered in Section~\ref{casestudies}.

We also assess the effect of different mass and radius measurements in the literature for a given planet. For this purpose, we consider the planet TOI-732~c for which such measurements are available from three sources as shown in Table~\ref{table:targets}. In Figure~\ref{fig:depths732} we show the range of ocean depths obtained using the mass and radius values from each of the three sources. The \citet{Cloutier2020} (hereafter referred to as "C20") range is the same as in Figure~\ref{fig:depths}. If we instead consider the \citet{Nowak2020} values (hereafter referred to as "N20") we find that somewhat deeper oceans are possible. The more precise measurements of N20 compared to C20 result in a smaller $\mathrm{log}\ g$ range at lower values, allowing larger depths. Conversely, the similar maximum $\mathrm{log}\ g$ for \citet{Bonfanti2024} ("B23") and C20 result in a similar upper limit for ocean depth. The improved precision of the B23 measurements again results in a narrower $\mathrm{log}\ g$ range and hence a smaller range of ocean depths. We note that the equilibrium temperature is not varied between the three examples, given the similar estimates between the three studies and the surface temperature being unconstrained. For example,  $T_\mathrm{eq}$ is $305-363$~K for N20 compared to $297-353$~K for C20, for $0\leq A_\mathrm{B}\leq0.5$.

\subsection{Case Studies} \label{casestudies}

We now explore in detail the possible range of ocean depths and atmospheric mass fractions for five promising Hycean candidates with upcoming JWST observations. The properties of these planets and their host stars are listed in Table~\ref{table:targets}. The planets include K2-18~b, TOI-270~d, TOI-1468~c, TOI-732~c and LHS 1140 b. These planets are shown against the Hycean mass-radius ($M$-$R$) plane from \citet{Madhusudhan2021} in Figure~\ref{fig:MR}. We estimate the maximum mass fraction in the H/He envelope for these planets for two end-member cases -- as a Hycean world, and as a rocky planet with a deep H$_2$-rich atmosphere and no H$_2$O. In the latter case, the maximum density interior of a pure Fe core is assumed in order to obtain the upper limit for H/He mass fraction. We note this is purely an upper limit, as a $100\%$ Fe core is unrealistic by planet formation scenarios. The upper H/He limit has implications for planet formation studies, which is discussed further in Section \ref{HHe}.

For each planet, we consider equilibrium temperatures assuming Bond albedos in the range $0\leq A_\mathrm{B}\leq0.5$. This is motivated by the values of \citet{Madhusudhan2021} -- the maximum $A_{\mathrm{B}}=0.5$ is used for their calculation of the inner habitable zone boundary for Hycean planets, motivated by \citet{Selsis2007,Yang2013,DePater2010}. The equilibrium temperature is calculated by

\begin{equation}
    T_{\mathrm{eq}} = T_{\star} \left[ \frac{R_{\star}^{2}}{2a^{2}}  \left(1-A_{\mathrm{B}}\right) \left(1-f_{\mathrm{r}}\right)\right]^{\frac{1}{4}}
    \label{eqn:equilibriumtemp}
\end{equation}

where $a$ is the orbital semi-major axis, $R_\star$ and $T_\star$ are the stellar radius and effective temperature respectively, and $f_\mathrm{r}$ is the fraction of incident radiation that is redistributed to the nightside. In this work we assume that this redistribution is efficient, adopting $f_\mathrm{r}=0.5$ for uniform day-night energy redistribution, giving a uniform equilibrium temperature across the planet. The Hycean candidates we consider here are expected to be tidally locked. Depending on the efficiency of day-night redistribution, there could be differences between the temperature of the day and night sides. Future work would be needed to study the effects of inefficient day-night redistribution on their interiors and oceans, including the case of Dark Hyceans \citep{Madhusudhan2021}.

In Section~\ref{Depths} we assumed constant gravity in the atmosphere, to demonstrate how initial estimates of the possible ocean depths could be obtained via Figure~\ref{fig:depths}. As described in Section~\ref{Depths}, the depths in Figure~\ref{fig:depths} were calculated assuming a fixed $P_\mathrm{HHB}$ of $100$ bar. In the subsequent sections, we discuss individual cases with gravity varying in the atmosphere. We evaluate the internal structure model across the full range of possible layer mass fractions to identify the compositions that satisfy the $1\sigma$ range of $M_\mathrm{p}$ and $R_\mathrm{p}$, including compositions that allow for Hycean conditions. The $P_\mathrm{HHB}$ and $T_\mathrm{HHB}$ are not fixed, with the allowed range for Hycean solutions spanning $273 \leq T_\mathrm{HHB} \leq400$ K and $1 \leq P_\mathrm{HHB} \leq 1000$ bar. As described in Section \ref{PTprofiles}, the atmospheric $P$-$T$ profiles are assumed to be isothermal/adiabatic profiles, with $P_\mathrm{rc}=100$ bar. $T_0$ is taken to span the $T_\mathrm{eq}$ range between $A_\mathrm{B}=0$ and $A_\mathrm{B}=0.5$, as outlined above. $P_0$ is taken to be the reference pressure for $R_\mathrm{p}$ where available, and otherwise we adopt the reference pressure of $0.05$ bar, following \citet{Madhusudhan2020}. We consider end-member core compositions of Earth-like ($33\%$ Fe) and pure Fe.

\subsubsection{TOI-270 d}\label{TOI270d}

TOI-270~d is a sub-Neptune discovered with TESS \citep{Gunther2019}, with RV follow-ups with ESPRESSO \citep{VanEylen2021}. The TOI-270 system contains three transiting planets orbiting an M3V host star with near-mean motion resonance, allowing recent transit-timing variations to be detected \citep{Kaye2022}. Both TOI-270~c and TOI-270~d are Hycean candidates \citep{Madhusudhan2021}. TOI-270~d orbits at $0.07$ au and has $M_\mathrm{p}=4.20\pm0.16\ \mathrm{M_\oplus}$ \citep{Kaye2022} and $R_\mathrm{p}=2.19\pm0.07\ \mathrm{R_\oplus}$ \citep{MikalEvans2023}. TOI-270~d is hottest of the Hycean candidates we consider in this paper, with $T_0$ taken to be $326-387$~K (this is the $T_\mathrm{eq}$ range for $0\leq A_\mathrm{B}\leq0.5$, as outlined above). This planet was recently the subject of an atmospheric study with HST, which suggested an H$_2$-rich atmosphere with H$_2$O absorption \citep{MikalEvans2023}. We take $P_0$ as $0.0912$ bar, which is the reference pressure for $R_\mathrm{p}$ \citep{MikalEvans2023}.  

We find the possible Hycean ocean depths for TOI-270~d span $\sim$$200-500$~km, $\sim$$50-140$ times the average depth of Earth's ocean at $3.7$ km \citep{Charette2010} ($\sim$$20-50$ times the deepest part at $\sim$$11$ km \citep{Gardner2014}). As discussed in Section \ref{Depths}, the maximal ocean depth is achieved with the minimum surface gravity and maximal surface temperature. In Figure~\ref{fig:interiors} we show an example interior for TOI-270~d with a maximal ocean depth of $500$~km. This case has the lower bound $M_\mathrm{p}=4.04\ \mathrm{M_\oplus}$, with $R_\mathrm{p}=2.22\ \mathrm{R_\oplus}$. The H$_2$O mass fraction is at its maximum $x_{\mathrm{H_2O}}=90\%$, with $x_\mathrm{H/He}=0.011\%$ and the remaining mass in an Earth-like core. The surface temperature lies at the maximum of $400$~K, at pressure $112$ bar -- more than $100$ times the surface pressure on Earth. Conversely, an example interior at the lower end of ocean depths has $x_{\mathrm{H_2O}}=86\%$, $x_{\mathrm{H/He}}=0.0072\%$ and the remainder in an Earth-like core, for $M_\mathrm{p}=4.264\ \mathrm{M_\oplus}$ and $R_\mathrm{p}=2.13\ \mathrm{R_\oplus}$. The depth in this case is $221$ km, with the HHB at the minimum temperature, $326$~K, and $90$ bar. We note that the interiors able to achieve a given ocean depth are degenerate. The minimum mass fraction of H$_2$O required for TOI-270~d to be a Hycean world is found to be $68\%$. This occurs for a low density interior, with maximum surface temperature $400$~K and an Earth-like core composition. These constraints on the H$_2$O content of sub-Neptunes can be useful for testing planet formation scenarios \citep[e.g.][]{Bitsch2021}.

We require a H/He mass fraction of $\lesssim0.0195\%$ for TOI-270~d to be a Hycean world. This limit corresponds to the maximal H$_2$O mass fraction of $90\%$ and an Earth-like core of $9.9815\%$, with the isothermal temperature in the atmosphere $T_0=326$~K. The HHB in this case is $197$ bar and at the maximum of $400$ K, as an increase in H/He mass fraction would increase the temperature at the surface beyond the habitable range. Conversely, a minimum H/He mass fraction of $1.7\times10^{-6}$ is required for Hycean conditions. This limit occurs for the maximum $T_0=387$~K and $90\%$ H$_2$O and an Earth-like core, giving an HHB at $20$ bar and $387$ K. Lower H/He mass fractions produce an $R_\mathrm{p}$ below the lower $1\sigma$ limit.

If we consider our other end-member case of a rocky planet with an H/He envelope, we find a maximum envelope fraction of $6.1\%$. This is achieved using the lower $T_0$ of $326$~K and a pure Fe core, with no H$_2$O. An Earth-like core would reduce the maximum H/He fraction to $3.7\%$, for the same $P$-$T$ profile.

We have adopted $P_\mathrm{rc}=100$ bar for the H/He envelope. As shown by \citet{Nixon2021}, the choice of $P_\mathrm{rc}$ affects the maximum H/He fraction that allows for an HHB within the liquid phase. A lower $P_\mathrm{rc}$ results in lower permitted H/He fractions, as the atmospheric temperature increases higher in the atmosphere. If we adopt $P_\mathrm{rc}=10$ bar and $P_\mathrm{rc}=1000$ bar, we find $\sim0.0017\%$ and $\sim0.0810\%$ as the maximum H/He fractions for Hycean conditions respectively. The overall maximum H/He fraction then varies from $3.4-8.3\%$ for $P_\mathrm{rc}$ from $10-1000$ bar. A higher $P_\mathrm{rc}$ would reduce the range of ocean depths possible, as this restricts the range of $T_\mathrm{HHB}$ to below the maximum possible of $400$~K. Reducing $P_\mathrm{rc}$ will not significantly affect the range of ocean depths, as we can already reach the maximum $T_\mathrm{HHB}=400$ K with $P_\mathrm{rc}=100$ bar.

\subsubsection{TOI-1468 c}\label{TOI1468c}

TOI-1468~c is a recently discovered sub-Neptune \citep{Chaturvedi2022} orbiting its M3V host star at $0.0859$ au, along with the closer-in, rocky TOI-1468~b. TOI-1468~c has $M_\mathrm{p}=6.64_{-0.68}^{+0.67}\  \mathrm{M_\oplus}$ and $R_\mathrm{p}=2.064\pm0.044\ \mathrm{R_\oplus}$ \citep{Chaturvedi2022}, with $T_0$ ranging from $284-338$~K for our $A_\mathrm{B}$ range. For TOI-1468~c and the remaining planets, we adopt $P_0=0.05$~bar, which is the median reference pressure from \citet{Madhusudhan2020} for K2-18~b. 

The possible ocean depths for TOI-1468~c are found to span $\sim 60$-$310$~km. An example internal structure that facilitates a maximal depth of $309$~km is shown in Figure~\ref{fig:interiors}. This solution has $x_{\mathrm{H/He}}=0.0125\%$, $x_{\mathrm{H_2O}}=63\%$ and a pure Fe core, for lower bound $M_\mathrm{p}=5.97\ \mathrm{M_\oplus}$ and upper bound $R_\mathrm{p}=2.108\ \mathrm{R_\oplus}$. The surface temperature lies close to the habitable maximum, at $397$~K, with an envelope $T_0$ of $284$~K. The surface pressure is $\sim300$ times Earth's surface pressure, at $307$ bar, while the base of the ocean lies at $6.0\times10^{4}$ bar. This case also represents the solution with maximum envelope mass fraction while maintaining Hycean conditions. Overall we find a maximum permitted H/He fraction of $4.3\%$ for non-Hycean conditions -- this is lower than for TOI-270~d due to the higher bulk density of TOI-1468~c. For an Earth-like core, as opposed to a pure Fe core, the maximum H/He mass fraction is $1.8\%$.

Unlike TOI-270~d, TOI-1468~c is not permitted to have $90\%$ H$_2$O, again due to its higher bulk density, with the maximum $x_{\mathrm{H_2O}}$ found to be $\sim$$75\%$. The minimum mass fraction of H$_2$O required for TOI-1468~c to be a Hycean world is found to be $24\%$. Using this minimum mass fraction of H$_2$O, we find the lower bound for high-pressure ice thickness on TOI-1468~c (while maintaining Hycean conditions) to be $0.42\ \mathrm{R_\oplus}$, or $\sim$$2700$ km. Even in this case, the ocean is still very deep, at $230$ km. The thickness of the high-pressure ice layer on Hycean worlds could have implications for their habitability, which is discussed further in Section \ref{Habitability}.

\begin{figure}
    \centering
    \includegraphics[width=\columnwidth]{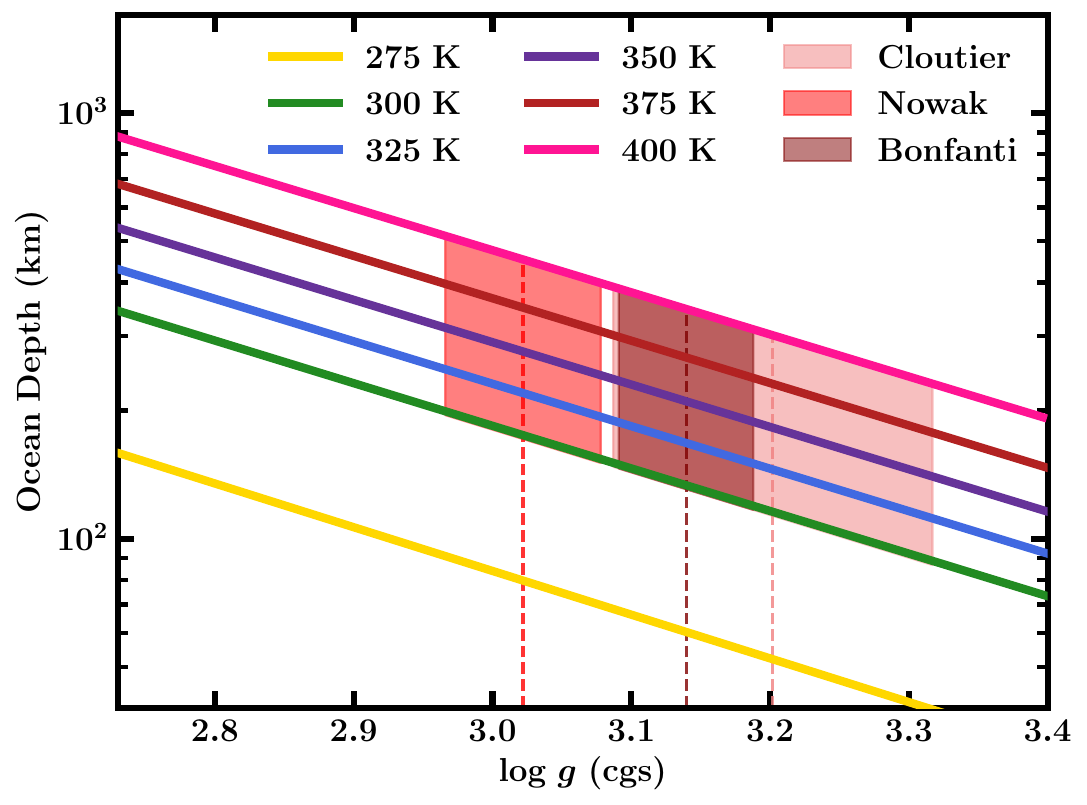}
    \caption{Ocean depth against surface gravity for a range of habitable surface temperatures. The three possible ranges for TOI-732~c for the alternative mass and radius measurements \citep{Cloutier2020,Nowak2020,Bonfanti2024} are shown by the shaded regions.}
    \label{fig:depths732}
\end{figure}

\begin{figure}
    \centering
    \includegraphics[width=\columnwidth]{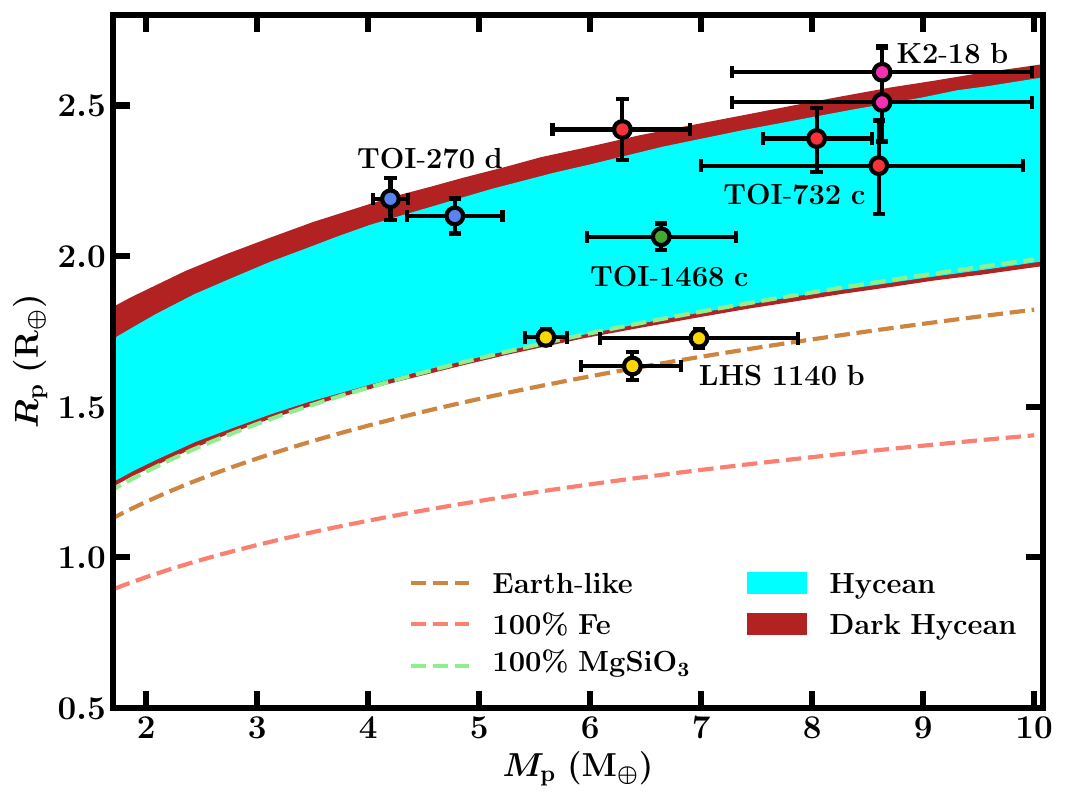}
    \caption{Hycean mass-radius plane from \citet{Madhusudhan2021}. The Hycean candidates discussed are also shown -- different mass/radius measurements for each planet are shown in corresponding colours.}
    \label{fig:MR}
\end{figure}

\subsubsection{TOI-732 c}\label{TOI732c}

TOI-732~c, or LTT 3780~c, is a sub-Neptune orbiting an M4V host star discovered by TESS, with independent follow-ups with HARPS \& HARPS-N \citep{Cloutier2019} and CARMENES \citep{Nowak2020}. Most recently, combined analysis of TESS, CHEOPS and ground-based light curves was carried out by \citet{Bonfanti2024}, along with RV analysis with MAROON-X. This system also contains a super-Earth, TOI-732~b, making this system an interesting test of radius valley formation around M dwarfs \citep{Cloutier2019}. TOI-732~c orbits its host star at $0.08$ au \citep{Cloutier2019,Nowak2020}, giving equilibrium temperatures in the range $284$-$353$~K for our $A_\mathrm{B}$ range. Multiple measurements of mass and radius are reported for TOI-732~c, with the masses varying significantly. \citet{Cloutier2020} find $M_\mathrm{p}=8.60^{+1.60}_{-1.30}\ \mathrm{M_\oplus}$ and $R_\mathrm{p}=2.30^{+0.16}_{-0.15}\ \mathrm{R_\oplus}$ while \citet{Nowak2020} find $M_\mathrm{p}=6.29^{+0.63}_{-0.61}\ \mathrm{M_\oplus}$ and $R_\mathrm{p}=2.42\pm0.10\ \mathrm{R_\oplus}$. The most recent mass measurement by \citet{Bonfanti2024} is in best agreement with \citet{Cloutier2020}, at $M_\mathrm{p}=8.04^{+0.50}_{-0.48}\ \mathrm{M_\oplus}$, with $R_\mathrm{p}=2.39^{+0.10}_{-0.11}\ \mathrm{R_\oplus}$. These are shown by the blue points in Figure~\ref{fig:MR}. In the following calculations and Figure~\ref{fig:depths}, we adopt the \citet{Cloutier2020} values (hereafter referred to as "C20").

\begin{figure*}
    \centering
    \includegraphics[width=0.6\columnwidth]{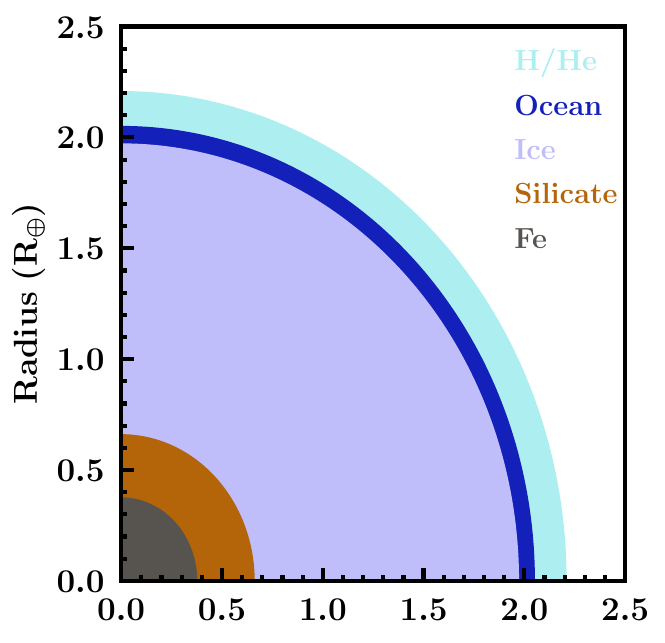}
    \includegraphics[width=0.6\columnwidth]{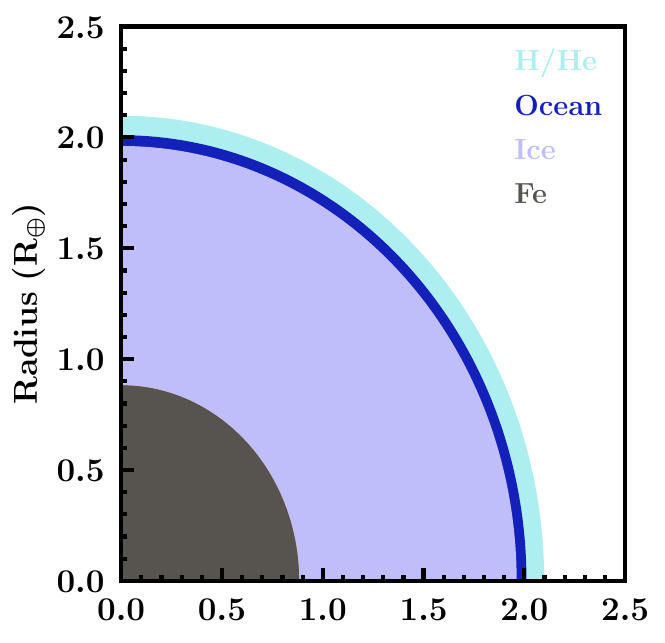}
    \includegraphics[width=0.6\columnwidth]{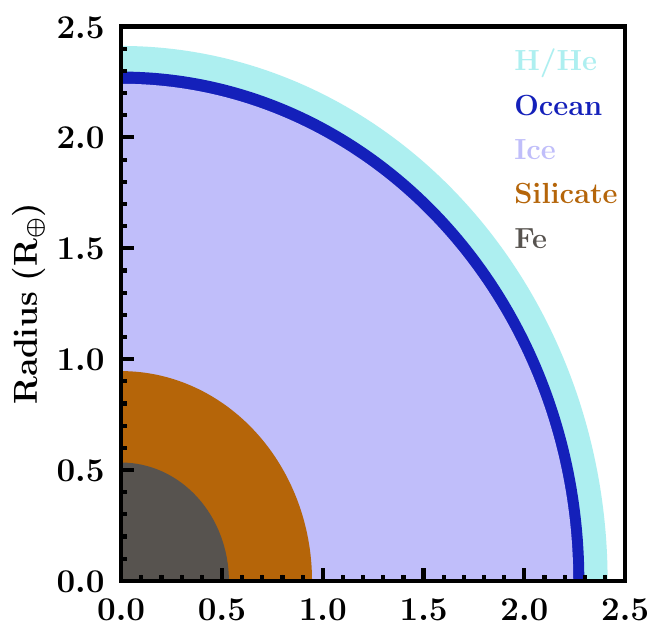}
    \caption{Possible interiors of candidate Hycean worlds. a: TOI-270~d, with ocean depth $500$ km. b: TOI-1468~c, with ocean depth $309$ km. This case has a pure Fe core, which is likely unrealistic based on planet formation mechanisms. c: TOI-732~c, with ocean depth $350$ km.}
    \label{fig:interiors}
\end{figure*}

We find possible Hycean ocean depths to range from $\sim$$90-360$ km. An example internal structure with depth $350$ km is shown in Figure~\ref{fig:interiors}. This case, with $M_\mathrm{p}=7.10\ \mathrm{M_\oplus}$ and $R_\mathrm{p}=2.41\ \mathrm{R_\oplus}$, has $x_{\mathrm{H/He}}=0.0132\%$, $x_{\mathrm{H_2O}}=80\%$ and the remainder in an Earth-like core. The surface temperature lies close to the maximum value at $399$~K, with a pressure of $270$ bar, for $T_0=297$~K. We find a maximum H/He fraction of $0.0142\%$ for TOI-732~c under Hycean conditions. As for TOI-270~d, this is achieved with a maximal H$_2$O fraction of $90\%$ with $9.986\%$ in a pure Fe core, and a minimal $T_0$, here $297$~K. The overall maximum H/He fraction for non-Hycean conditions is found to be $6.8\%$, using the lower $T_0$ of $297$~K and a pure Fe core. A maximum envelope fraction of $3.8\%$ is permitted for an Earth-like core. TOI-732~c has the largest range of permitted H$_2$O fractions of the planets considered, due to the larger uncertainty in its mass. The minimum mass fraction of H$_2$O required for TOI-732~c to be a Hycean world is found to be $22\%$, and the interior in Figure~\ref{fig:interiors} demonstrates solutions are possible up to $90\%$ H$_2$O.

As shown in Section \ref{Depths}, assuming constant gravity in the atmosphere, different mass measurements for TOI-732~c can affect the possible range of ocean depths. Adopting the \citet{Nowak2020} values ("N20") in our full evaluation of the model (assuming the same range for $T_\mathrm{eq}$ as used for C20), the maximum depth is found to be $415$ km, compared to $350$ km for the C20 case. This difference is smaller than Figure~\ref{fig:depths732} would suggest, due to the surface conditions permitted by the results of possible compositions. The maximum H/He fraction is found to be higher than for C20, at $8.2\%$. For Hycean conditions, the maximum H/He fraction for N20 values is only slightly higher, at $0.0186\%$.

\subsubsection{K2-18 b}\label{K218b}

We revisit K2-18~b, a well-studied sub-Neptune \citep[e.g.][]{Benneke2017,Cloutier2019,Benneke2019,Tsiaras2019,Madhusudhan2020,Blain2021,Madhusudhan2023b} and the first Hycean candidate \citep{Madhusudhan2020,Madhusudhan2021}. K2-18~b orbits it M3V host star at $0.15$ au, and has $M_{\mathrm{p}}=8.63\pm1.35\ \mathrm{M_\oplus}$ \citep{Cloutier2019} and $R_{\mathrm{p}}=2.610\pm0.087\ \mathrm{R_\oplus}$ \citep{Benneke2019}, or $R_{\mathrm{p}}=2.51^{+0.13}_{-0.18}\ \mathrm{R_\oplus}$ \citep{Hardegree2020}. We use the former $R_\mathrm{p}$ value for consistency with previous studies \citep{Madhusudhan2020}. As listed in Table~\ref{table:targets}, $T_{\mathrm{eq}}$ varies between $250$~K and $297$~K at $A_\mathrm{B}=0.5$ and $0$ respectively.

We first adopt PT3 from \citet{Madhusudhan2023a}, shown in Figure~\ref{fig:envelopePT}. This profile was generated via self-consistent atmospheric modelling with the GENESIS framework \citep{Gandhi2017,Piette2020} -- see \citet{Madhusudhan2023a} for a full description. This profile was calculated to a pressure of $100$ bar at which it has temperature $387$~K, and we extend to $P>100$ bar with an adiabatic profile. We note that this profile is different to those used by \citet{Madhusudhan2020} in their internal structure modelling of K2-18~b. They adopt two atmospheric $P$-$T$ profiles, also generated with GENESIS, which vary in assumptions, including internal temperature. These have $T_\mathrm{int}=25$ K and $T_\mathrm{int}=50$ K, while PT3 from \citet{Madhusudhan2023a}, used in this work, has $T_\mathrm{int}=30$ K; see \citet{Madhusudhan2020} and \citet{Madhusudhan2023a} for full descriptions of the assumptions made.

Using PT3, we find the possible ocean depths to be $140-350$ km. The $T_\mathrm{HHB}$ range possible using PT3 is limited, since at the minimum $P_\mathrm{HHB}$ of $1$~bar, PT3 is at $316$~K. To be a Hycean world, we find K2-18 b requires a H/He fraction of $\lesssim 0.0052\%$. \citet{Madhusudhan2020} find permitted envelope fractions of $0.006\%$ for an interior solution with a liquid surface, consistent with our result. Overall we obtain a maximum H/He fraction of $8.1\%$ for non-Hycean conditions, with a pure Fe core. In comparison, \citet{Madhusudhan2020} found $6.2\%$ for the maximum H/He mass fraction. For an Earth-like core composition, we find a lower maximum H/He fraction of $4.9\%$.

If we instead adopt isothermal/adiabatic profiles as we have for the other case studies, we expect the aforementioned importance of the atmospheric $P$-$T$ profile to affect the results. At pressures $\gtrsim 1$ bar PT3 is at higher temperatures than the isothermal/adiabatic cases at $T_0=250$~K and $T_0=297$~K. Therefore, we find that a larger H/He fraction is permitted for the isothermal/adiabatic case. We adopt $T_0=250$~K, which is consistent with the isothermal profile found in the photosphere via the retrieval carried out by \citet{Madhusudhan2023b}. As with previous case studies, we adopt $P_\mathrm{rc}=100$~bar. We find a maximum H/He fraction for Hycean conditions of $0.022\%$ for this $T_0=250$~K and $T_\mathrm{HHB}=400$~K. In this case, the range of ocean depths is also affected as lower values of $T_\mathrm{HHB}$ are accessible with this $P$-$T$ profile. For example, an interior with upper bound $M_\mathrm{p}$, $x_{\mathrm{H/He}}=0.0045\%$, $x_\mathrm{H_2O}=86\%$ and the remainder an Earth-like core results in a nearly minimal depth of $54$~km, at $T_\mathrm{HHB}=277$~K and $P_\mathrm{HHB}=139$~bar. The pressure at the base of the ocean in this case is $1.0\times10^{4}$ bar, compared to $\sim$$6\times10^{4}$ bar for a $400$~K surface. The upper depth limit is not significantly affected by a variation in atmospheric $P$-$T$ profile as the upper limit for $T_\mathrm{HHB}$ and hence ocean base pressure is unchanged. Reducing $T_0$ can allow for even larger H/He mass fractions while maintaining Hycean conditions. For instance, adopting $T_0=200$~K results in a maximum H/He mass fraction of $0.046\%$ for Hycean conditions, about twice that for $T_0=250$~K. In this case, the HHB lies close to the maximum for habitable conditions, at $\sim$$1000$ bar.

We note that the depth estimates depend strongly on the HHB conditions, which in turn depend on the temperature structure and atmospheric composition. The non-detection of H$_2$O in the photosphere of K2-18~b may be consistent with the presence of a tropospheric cold trap, with the temperature and H$_2$O abundance potentially higher in the lower atmosphere \citep{Madhusudhan2023b}. However, it is difficult to accurately estimate the composition and temperature structure of the dayside atmosphere, and the corresponding surface temperature and pressure, based on observed photospheric properties at the day/night terminator using transmission spectroscopy. We have, therefore, considered a wide range for the $T_\mathrm{HHB}$ for K2-18~b, similar to the other planets in this study, to consistently explore the full range of possibilities.

Across all the $P$-$T$ profiles considered for K2-18~b the ocean depths span $\sim$$50-350$~km. However, the non-detection of H$_2$O in K2-18~b \citep{Madhusudhan2023b} may limit the surface temperatures to well below $400$~K. We therefore evaluate the possible ocean depths for a lower maximum $T_\mathrm{HHB}$ of $340$~K. Adopting PT3 as the atmospheric profile for this case, we find a narrow ocean depth range of $140-180$~km. Adopting an isothermal/adiabatic profile at $T_0=250$~K and $P_\mathrm{rc}=100$~bar, we find a range of $\sim$$50-180$ km. As mentioned previously, given the same maximum $T_\mathrm{HHB}$, the upper depth limit is not significantly affected by variations in atmospheric $P$-$T$ profile. However, the lower depth limit is affected as a lower $T_\mathrm{HHB}$, down to $273$~K, is possible for the latter $P$-$T$ profile. Similarly, considering an even lower maximum $T_\mathrm{HHB}$ further decreases the maximum ocean depth possible, e.g. to $\sim$$50-120$ km for $300$~K maximal $T_\mathrm{HHB}$ for the isothermal/adiabatic profile. Reducing the maximum $T_\mathrm{HHB}$ also decreases the maximum possible H/He mass fractions for Hycean conditions. For instance, using the isothermal/adiabatic profile the maximum H/He fraction is approximately halved for a maximum $T_\mathrm{HHB}$ of $340$~K compared to  $400$~K, at $\sim$$0.011\%$ vs $0.022\%$.

\subsubsection{LHS 1140 b} \label{LHS1140}

Finally, we discuss the case of LHS 1140~b \citep{Dittmann2017,Ment2019,Lillo2020,Cadieux2024}. This planet has a low equilibrium temperature of $235$~K for $A_\mathrm{B}=0$ ($198$~K for $A_\mathrm{B}=0.5$), orbiting its M4.5V host star at $0.0936$~au \citep{Ment2019}. Until recently, LHS 1140~b would only have been considered a Hycean candidate for H$_2$O mass fractions below $10\%$ \citep{Madhusudhan2021}. However, the bulk properties of planet have been revised recently with  $M_\mathrm{p}=5.60\pm0.19$ $\mathrm{M_\oplus}$ and $R_\mathrm{p}=1.730\pm0.025$ $\mathrm{R_\oplus}$ \citep{Cadieux2024}. These latest measurements place LHS 1140~b within the Hycean $M$-$R$ plane, and is close to the upper density limit for Hycean candidates, as shown in Figure~\ref{fig:MR} along with the previously reported measurements. For reference, \citet{Ment2019} reported values $M_\mathrm{p}=6.98\pm0.89$ $\mathrm{M_\oplus}$ and $R_\mathrm{p}=1.727\pm0.032$ $\mathrm{R_\oplus}$, placing this planet just outside the upper density limit for Hycean candidates. On the other hand, \citet{Lillo2020} reported  $M_\mathrm{p}=6.38^{+0.46}_{-0.44}$ $\mathrm{M_\oplus}$ and $R_\mathrm{p}=1.635\pm0.046$ $\mathrm{R_\oplus}$. \citet{Madhusudhan2021} explain that the lower boundary could be closer to the $M$–$R$ curve for an Earth-like composition if lower H$_2$O mass fractions were permitted -- the minimum is assumed to be $10\%$, as in this study. LHS~1140~b therefore represents an end-member case, in density and temperature, for Hycean candidates.

We first use the \citet{Ment2019} values, as the most conservative case. We adopt the equilibrium temperature $T_0=235$~K and a $P$-$T$ profile in the envelope with $P_\mathrm{rc}$ at $100$ bar. Even for this conservative case, large ocean depths are possible. For instance, fractions of $x_{\mathrm{H_2O}}=30\%$, $x_{\mathrm{H/He}}=8.0\times10^{-5}$ and remainder pure Fe results in a $145$ km ocean. The surface temperature is $360$~K in this case. Therefore, LHS 1140~b demonstrates that even planets on the extreme boundaries of being candidate Hycean worlds under certain conditions could host 100s of km deep oceans. If we instead use the latest \citet{Cadieux2024} values for $M_\mathrm{p}$ and $R_\mathrm{p}$, giving a lower surface gravity, we find significantly deeper oceans are possible for LHS 1140~b, up to $\sim$$300$ km. For example, mass fractions of $x_{\mathrm{H_2O}}=50\%$, $x_{\mathrm{H/He}}=0.021\%$ and remainder in pure Fe can result in a $290$ km ocean, for $T_\mathrm{HHB}=400$ K. This is calculated adopting the same atmospheric $P$-$T$ profile as used with the \citet{Ment2019} values. If we instead assume an atmospheric profile with the lower $T_0=198$~K and $P_\mathrm{rc}=100$ bar, we obtain the maximum H/He fraction for Hycean conditions to be $0.036\%$. Conversely, for non-Hycean conditions, the maximum H/He fraction is found to be $4.2\%$, using the same $P$-$T$ profile and the \citet{Cadieux2024} values, with a pure Fe core. For an Earth-like core, the maximum H/He fraction is reduced to $1.7\%$. 

\subsection{Envelope Mass Fractions on Hycean Worlds}\label{HHeHycean}

The requirements of a habitable ocean places limits on the possible mass fraction of the H/He envelope in a given Hycean world. The envelope mass fraction depends on the temperature structure and the HHB conditions at the ocean surface. Generally, lower atmospheric temperatures allow for higher envelope mass fractions. For example, higher $P_\mathrm{rc}$ and lower $T_0$ allow for higher envelope mass fractions, and vice versa. Across our case studies and the assumptions considered in this work we find the maximum envelope mass fractions admissible for Hycean conditions to be $\sim$$10^{-3}$. Higher envelope mass fractions correspond to cooler atmospheric temperatures and higher surface pressures. Improved constraints on H/He mass fraction can be obtained via better constraints on the atmospheric temperature structure. Atmospheric observations, for instance with the JWST, are essential for obtaining these improved constraints \citep[e.g.][]{MikalEvans2023,Madhusudhan2023b}. See Section \ref{Observations} for further discussion of upcoming observations.

Presently, the formation mechanisms of Hycean worlds with such envelope mass fractions have not been investigated in detail. More generally, several mechanisms have been proposed for the formation and evolution of sub-Neptune planets, especially with the aim of explaining the radius valley \citep{Fulton2017,Fulton2018}. These mechanisms include processes where thick primordial H$_2$-rich envelopes are depleted via photoevaporative mass loss \citep[e.g.][]{Owen2017} and/or core-powered mass loss \citep[e.g.][]{Gupta2019}, as well as processes involving outgassing of H$_2$ from the interiors \citep[e.g.][]{ElkinsTanton2008}. Other mechanisms suggest the preponderance of sub-Neptunes with water-rich interiors and envelopes of varied compositions \citep[e.g.][]{Zeng2019,Venturini2020,Izidoro2022}, which could include Hycean worlds. Considering photoevaporative and core-powered stripping of the envelopes of sub-Neptunes with 1:1 silicate-to-ice ratios, \citet{Rogers2023} find envelope mass fractions of $\gtrsim10^{-3}$ can be retained, at $T_\mathrm{eq}=300$~K, after $5$~Gyr of photoevaporative evolution. \citet{Izidoro2022} find that water-rich sub-Neptunes with H$_2$-rich atmospheres, which could include Hyceans, can be formed via gas-driven migration models, both with and without the inclusion of photoevaporative mass loss. However, the possible envelope mass fractions were not constrained in this study -- a fixed fraction of $0.3\%$ was assumed, based on \citet{Zeng2019}. Such mechanisms could have varying implications for formation of Hycean planets.

We note that an H$_2$-rich envelope mass fraction of $\sim$$10^{-4}-10^{-3}$ is $10^{2}-10^{3}$$\times$ larger than that of the Earth, albeit with a lower mean molecular weight. These required mass fractions open a new avenue for investigating the origins of Hycean worlds. In principle, these mass fractions are at the limit of what could be retained by mass loss mechanisms in temperate sub-Neptunes based on recent studies \citep[e.g.][]{Owen2017,Gupta2019,Rogers2023}. On the other hand, whether outgassing \citep[e.g.][]{ElkinsTanton2008} or other atmosphere/ocean exchange processes can result in these mass fractions remains to be seen.

\subsection{Maximum Envelope Mass Fractions}\label{HHe}

We have additionally placed constraints on the maximum envelope mass fraction for each of the sub-Neptunes we consider. The extreme case assumes a pure Fe interior with no H$_2$O, with our standard $P_\mathrm{rc}=100$ bar in the envelope. We find that the upper envelope fractions are all within $\sim4$-$8\%$. This is similar to the $\lesssim 7\%$ found by \citet{Valencia2013} for the well-studied GJ 1214 b. The maximum fraction found by \citet{Madhusudhan2020} for K2-18b is also similar, at $\sim6\%$. For an Earth-like core composition, we find the maximum H/He envelope mass fractions to span $\sim$$2$-$5\%$.

The upper limit for envelope mass fraction has implications for planet formation and evolution scenarios in the sub-Neptune regime. Mechanisms of atmospheric mass loss, including both photoevaporative \citep[e.g.][]{Owen2013,Owen2017,Rogers2021} and core-powered \citep[e.g.][]{Gupta2019,Gupta2020}, make predictions for permitted envelope mass fractions for sub-Neptunes. Determining these for a range of sub-Neptunes including those within and outside of the radius valley can help to test the predictions of these theories. For instance, the photoevaporative scenario \citep{Owen2017} predicts mass fractions of order $\sim$$1\%$ are typical for envelope-retaining sub-Neptunes. More recent studies also including core-powered mass loss suggest larger mass fractions up to $\sim$$10\%$ are possible, depending on the planet mass and radius \citep{Rogers2023}. Our range of derived H/He mass fractions are consistent with these estimates. 

\section{Summary and Discussion}

In this paper we investigate the range of conditions possible in the interiors of Hycean worlds, including their ocean depths, interior compositions and envelope mass fractions. Our results follow previous works on ocean depths in water-rich sub-Neptunes \citep{Noack2016,Nixon2021}, focusing specifically on Hycean conditions and several candidate Hycean worlds. Firstly, we find the expected range of ocean depths to extend from $10$s of km to $\sim$$1000$ km for Hycean worlds. The depth of a Hycean ocean is influenced by the surface conditions, specifically the temperature and gravity, and is hence sensitive to the planet mass, composition, and assumed temperature profile in the envelope. We constrain the possible ocean depths and compositions for a sample of five promising Hycean candidates with upcoming JWST data: TOI-270~d, TOI-1468~c, TOI-732~c, K2-18~b and LHS 1140~b. Secondly, we constrain the mass fractions of the possible H/He envelopes for all the candidates considered. Across the sample we find the maximum envelope fraction admissible for Hycean conditions to be $\sim$$10^{-3}$, for the atmospheric temperature structures explored in this work. Finally, we also constrain the maximum H/He envelope mass fraction for non-Hycean conditions, i.e. for the limiting case of a rocky core with a thick H/He envelope but no H$_2$O layer. The corresponding envelope mass fractions are found to span $\sim$$4-8\%$ across the sample.

Our results demonstrate the diverse conditions possible among Hycean worlds, and reinforce their possibility to host habitable conditions under vastly different circumstances to the Earth. The information we expect to gain on atmospheric composition with JWST will provide an indication if these planets could indeed be Hycean worlds and allow us to place better constraints on the nature of their interiors and oceans. With JWST, the detection of possible biomarkers in the atmospheres of Hycean candidates remains an exciting and potentially imminent prospect. The study by \citet{Madhusudhan2023b} of K2-18~b is an exciting first look into the capability of JWST to shed light on this regime.

\subsection{Habitability of Hycean Worlds}\label{Habitability}

\begin{figure}
    \centering
    \includegraphics[width=\columnwidth]{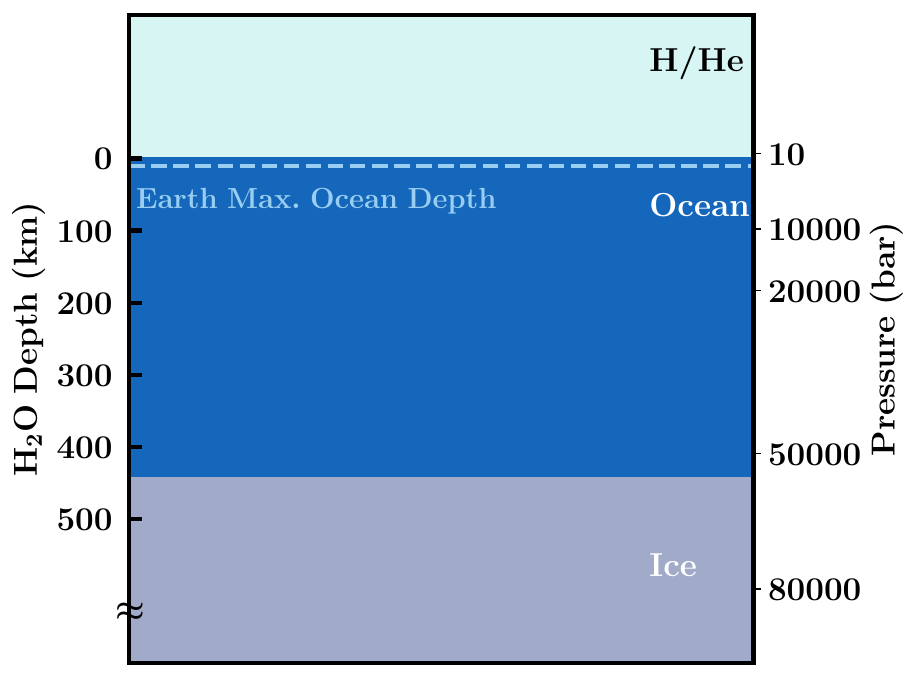}
    \caption{1-dimensional cross-section of a possible ocean of TOI-270~d. This corresponds to a case with ocean depth $443$ km. The maximum ocean depth on Earth is shown as a dashed line, at $11$ km.}
    \label{fig:ocean}
\end{figure}

The habitability of a Hycean world depends on a range of factors. In Figure~\ref{fig:ocean} we show an example of a possible ocean cross-section for the planet TOI-270~d. The maximum depth of Earth's ocean is shown by the dashed line, at $\sim$$11$ km \citep{Gardner2014}. Life in Earth's oceans spans the entire depth, down to pressures of $\sim$$1000$ bar \citep{Rothschild2001,Merino2019}. As we have shown in Section \ref{Results}, the depth of Hycean oceans can span hundreds of km down to pressures of $\sim$$10^{4}-10^{5}$ bar ($10^{9}-10^{10}$ Pa) where the transition to ice VI or VII occurs. In this example in Figure~\ref{fig:ocean}, the ocean base pressure is at $5.5\times10^{4}$ bar. Therefore, the majority of the ocean exists at temperatures and pressures greater than those known to host life on the Earth. We define the ``habitable depth'' as the depth at which we reach $P=1000$ bar and/or $T=400$ K. In this example in Figure~\ref{fig:ocean}, the habitable depth is $20$ km, where we reach $1000$ bar (at $394$ K), for an HHB at $\sim$$10$ bar and $387$ K. The habitable depth in this case is approximately double the deepest point of Earth's ocean. Assuming the same surface gravity and temperature, an HHB at lower pressure would result in a larger habitable depth, due to the dependence of depth on the change in pressure. However, changing the HHB pressure within our permitted range makes little difference to the overall ocean depth, which is evident from the rapid increase of pressure with depth shown in Figure~\ref{fig:ocean}. We should also note that it is unknown whether life could evolve to exist in conditions beyond these Earth-based pressures and temperatures, extending the potentially habitable portion of the ocean.

Hycean planets are generally expected to have high-pressure ice layers beneath their oceans. The presence of an icy mantle separating the ocean from the rocky core may have implications for the habitability of water-rich bodies \citep{Maruyama2013,Noack2016,Journaux2020b,Madhusudhan2023a}. For instance, the prevention of silicate weathering has been suggested to affect geochemical cycling \citep[e.g.][]{Kitzmann2015}. However, alternative mechanisms have already been proposed for ocean worlds with higher mean molecular weight atmospheres which can also possess high-pressure ice layers \citep{Kite2018,Levi2017,Ramirez2018}. The geochemical cycles that would occur in Hycean environments are unknown \citep{Madhusudhan2021} and future work is needed to investigate these possibilities. 

The implications of high-pressure ice layers have also been discussed in the context of ocean nutrient enrichment \citep[e.g.][]{Noack2016,Madhusudhan2023a}. The lack of contact between the ocean and mantle prevents the weathering of the seafloor that can enrich the ocean with nutrients needed for life. \citet{Madhusudhan2023a} suggest alternative ways to meet the chemical requirements for life in Hycean oceans, including atmospheric condensation and external delivery from asteroids/comets. However, transport across high-pressure ice layers has also been suggested as a possibility, based on a number of works studying convection in the high-pressure ices that could be present on water-rich exoplanets \citep{Choblet2017, Kalousova2018a, Kalousova2018b,Hernandez2022,Lebec2023,Madhusudhan2023a}, along with work on icy moons \citep[e.g.][]{Lingam2018,Journaux2020b}. The cases shown in Section \ref{Results} have high-pressure ice layers that can generally extend for $\gtrsim 1\ \mathrm{R_{\oplus}}$. As is intuitive, a smaller mass of H$_2$O results in a thinner high-pressure ice layer. For example, as discussed in Section \ref{TOI1468c}, for TOI-1468~c we find the ice can be as thin as $0.42\ \mathrm{R_\oplus}$ for an interior with the minimum permitted mass fraction of H$_2$O, while still maintaining an ocean depth of $189$ km. The thinner the ice layer, the smaller the distance required to transport nutrients across. However, the behaviour of high-pressure ices is largely unknown, including potential interactions between high-pressure ice and the underlying rock \citep[e.g.][]{Journaux2020b}, and future investigation is required to address these areas.

Our Hycean candidates orbit M dwarfs, which are generally more active than higher mass stars. Effects such as stellar wind and higher UV flux are able to erode a planet's atmosphere, which has the potential to make environments hostile to life \citep[e.g.][]{Shields2016}. However, the level at which a planet will be inhospitable is thought to be significantly dependent on the atmospheric composition and the level of host star activity \citep[e.g.][]{OMalleyJames2017}. As pointed out by \citet{Madhusudhan2021}, Hycean planets may be more habitable than terrestrial planets around M dwarfs due to their larger gravity and thicker atmospheres. Furthermore, for instance, TOI-270 has been shown by multiple studies \citep{Gunther2019,VanEylen2021,MikalEvans2023} to have low stellar activity levels. LTT 3780 (TOI-732) has also been found to be relatively inactive \citep{Nowak2020,Cloutier2020}, as has TOI-1468 \citep{Chaturvedi2022}. Therefore, these Hycean candidates may stand to be among the more promising candidates for life around M dwarfs. Nevertheless, habitability remains a complex topic and a rigorous assessment would need to be made on a planet-by-planet basis. 

\subsection{Mixed Envelopes} \label{MixedEnvelopes}

In this study we have not explicitly included mixed envelopes, with H$_2$O miscible in H/He. The difference in the radius obtained assuming a mixed vs an unmixed envelope is found to be less than the measured uncertainty in $R_\mathrm{p}$ of $\sim$$0.1\ \mathrm{R_\oplus}$ for the majority of the Hycean candidates. This is in agreement with \citet{Nixon2021} finding that $1\%$ H$_2$O mixed in a H/He atmosphere has a minimal effect on the radius. They adopted a $5\%$ envelope fraction, $T_0=500$ K, $P_0=0.1$ bar and $P_\mathrm{rc}=10$ bar, giving vapour or supercritical H$_2$O in the envelope to increase the likelihood of miscibility. It should be noted that for Hycean cases, the temperatures in the envelope are much less than considered by \citet{Nixon2021}, with a maximum of $400$ K. The envelope mass fractions of Hycean worlds are also $\sim$$10-100\times$ smaller than this. Therefore, the effect of the mixed envelope is expected to be even less. In their study of K2-18~b, \citet{Madhusudhan2020} find the median mixing ratio of H$_2$O to be $0.7$–$1.6\%$ across the models considered. With such mixing ratios, they find the radius difference is less than half the measured uncertainty when a mixed envelope is considered. However, the recent study with JWST found a non-detection of H$_2$O \citep{Madhusudhan2023b}, differing from the previous study with HST due to degeneracy with CH$_4$. The low H$_2$O mixing ratio implies the presence of a tropospheric cold trap resulting in condensation, with H$_2$O possibly more abundant below \citep{Madhusudhan2023a,Madhusudhan2023b}. 

Therefore, given the current level of uncertainty in radius measurements, incorporating the effect of a mixed envelope into our calculations would introduce an extra source of degeneracy in determining interior compositions, while not significantly affecting results for possible ocean depths. JWST observations of Hycean candidates and sub-Neptunes generally can better constrain their H$_2$O mixing ratios. For instance, for TOI-270~d \citet{MikalEvans2023} found with HST a $99\%$ credible upper limit of $30\%$ for the mixing ratio of H$_2$O, which ruled out a steam atmosphere. Given precise estimates, we can adopt representative mixed envelope EOSs in future studies of these planets. JWST observations will allow more precise H$_2$O abundance estimates for both K2-18~b and TOI-270~d, and observations of other candidate Hycean worlds will reveal the diversity in H$_2$O abundance. We note that as uncertainties in $R_\mathrm{p}$ measurements improve with next-generation facilities, it could become important to consider the effect of mixed envelopes.

\subsection{Future Directions for Internal Structure Modelling} \label{FutureIS}

A degenerate set of interior compositions can typically explain the bulk properties of an exoplanet in the sub-Neptune regime. Precise mass and radius measurements are therefore critical for internal structure modelling, as these can reduce the number of plausible solutions. This is evident from Figure~\ref{fig:depths732}, and from the variation in possible H/He mass fractions for TOI-732~c based on the available sources for $M_\mathrm{p}$ and $R_\mathrm{p}$ measurements. The other essential avenue is atmospheric observations, which can help to break the degeneracy -- this will be discussed further in Section \ref{Observations}.

There are a number of assumptions made in our internal structure model, including our adopted EOSs, that may be investigated in the future. We do not consider the effect of hydrated silicate/iron layers, which was investigated in recent studies \citep{Shah2021,Dorn2021}. However, the effect on the $M$-$R$ relation for solid rock in super-Earths was found to be smaller than the current typical precision of measurements \citep{Shah2021}. Furthermore, the behaviour of high-pressure ices is not fully known. This includes potential interactions between the high-pressure ices and underlying rock \citep{Vazan2022,Kovacevic2022}. \citet{Vazan2022} suggest that for the mass range $5-15\ \mathrm{M_{\oplus}}$, ice and rock can be mixed in the interiors of sub-Neptunes, but whether this occurs is dependent on the conditions at formation. 
We also note that the phase transitions and EOS of high-pressure ices are still very uncertain and require further study. Our H$_2$O EOS can be modified as new data becomes available. For instance, recent studies have revealed a new phase of ice VII known as ice VIIt \citep{Grande2022} which is suggested to have an effect on the $M$-$R$ relation comparable to observational uncertainties \citep{Huang2021}. 

We have adopted an adiabatic profile throughout the H$_2$O layer, as is commonly assumed in internal structure models for sub-Neptune-sized planets \citep[e.g.][]{Sotin2007,Nixon2021,Leleu2021}. However, the presence of thermal boundary layers in the interior would form barriers to convection and could affect the permitted compositions. The effect of these layers has been explored for Uranus and Neptune \citep[e.g.][]{Podolak2019}. 

\subsection{Observational Prospects} \label{Observations}

As we have shown in this paper, and has been discussed thoroughly in the literature \citep[e.g.][]{Rogers2010a,Valencia2013,Leleu2021,Nixon2021}, the bulk properties of a sub-Neptune are insufficient to place robust constraints on its composition, due to degenerate solutions. Atmospheric data are key for breaking these degeneracies. The first question for these planets will be establishing the presence or lack of an H$_2$-rich atmosphere. Atmospheric observations with HST and/or JWST have already confirmed H$_2$-rich atmospheres for K2-18~b \citep{Benneke2019,Tsiaras2019,Madhusudhan2023b} and TOI-270~d \citep{MikalEvans2023}. However, even if the presence of an H$_2$-rich atmosphere is established, Hycean worlds can be degenerate with sub-Neptunes with either an H$_2$-rich envelope and a solid rocky surface, or with a deep H$_2$-rich atmosphere that causes the surface to be too hot to sustain liquid H$_2$O. The essential step in diagnosing a Hycean world is thus establishing the presence of the surface ocean. This requires precise abundances for a number of different molecules and a comprehensive exploration of the possible chemical pathways on the planet given these abundances. The key molecules are H$_2$O, CH$_4$, NH$_3$, CO$_2$ and CO, in addition to any other hydrocarbons present \citep{Yu2021,Hu2021,Tsai2021,Madhusudhan2023a}. Figure~8 in \citet{Madhusudhan2023a} summarises the route to chemically diagnosing a Hycean world via these molecules. In the recent study by \citet{Madhusudhan2023b}, enhanced CH$_4$ and CO$_2$ were detected along with a lack of NH$_3$ in the observable atmosphere of K2-18~b, suggesting the presence of a surface ocean. This was carried out using JWST transmission spectra for one transit with each of NIRISS SOSS (Single Object Slitless Spectroscopy) and NIRSpec G395H. Additional upcoming JWST observations of K2-18~b, including one transit with MIRI LRS (Low Resolution Spectroscopy) via the same program (GO 2722) and multiple transits with NIRSpec G395H via GO 2372, can verify these detections. 

The planets TOI-270 d, TOI-1468 c and TOI-732 c are also scheduled for spectroscopic observations with JWST in Cycle 2. For each planet, at least three transits will be observed, one with each of NIRISS SOSS, NIRSpec G395H and MIRI LRS; for TOI-270 d, additional NIRISS and NIRSpec observations will be obtained. These will be observed in multiple programs (GO 3557, GTO 2759, GO 4098). As discussed above, the same combination of observations is also being carried out for K2-18 b in multiple programs (GO 2722 and GO 2372). The predicted uncertainties and long wavelength coverage are expected to allow robust detections of the key molecules required to diagnose a Hycean planet. The observations are hence expected to aid the distinction of a Hycean world from scenarios of a rocky planet with a thick H/He envelope, mini-Neptune or water world as outlined above, akin to the initial findings for K2-18 b \citep{Madhusudhan2023b}. LHS 1140 b has also been observed as part of Cycle 1 GO Program 2334, with a transit observed with each of NIRSpec G395H and G235H. 

Hycean worlds are promising candidates for biomarker detection due to their larger radii and higher temperatures compared to rocky planets. \citet{Madhusudhan2021} investigated the observability of biomarkers in the atmospheres of Hycean candidates K2-18 b, TOI-270 d and TOI-732 c, considering DMS, CS$_2$, CH$_3$Cl, OCS and N$_2$O as biomarkers. They predicted that the approved observations of K2-18 b in Cycle 1 would be sufficient to detect biomarkers in its atmosphere if present in the quantities considered. Potential evidence for DMS in the atmosphere of K2-18~b was suggested by \citet{Madhusudhan2023b}, though the abundance is not robustly constrained by the retrieval. They note the need for further theoretical exploration of atmospheric and interior processes when evaluating the viability of any possible biosignature. The additional observing time may better constrain the abundance in addition to potential detections of other species. 

The prospect of identifying Hycean worlds amongst the exoplanet population and potentially detecting signs of life on them has recently become a tangible possibility. There remains the exciting potential for life's existence on a planet vastly different to our own. Additional theoretical studies in the future could help further develop our understanding of Hycean worlds and their ability to support life. 

\section*{Acknowledgements}

FR and NM acknowledge support from the Science \& Technology Facilities Council (UKRI grant 2605554) towards the PhD studies of FR. We thank the anonymous reviewer for the careful review of our manuscript and helpful comments. FR thanks Matthew Nixon for model comparisons at the initial stage of the project.


\section*{Data Availability}

This work is theoretical and no new data is generated as a result. 
 


\bibliographystyle{mnras}
\bibliography{Sources} 

\begin{thebibliography}{}
\makeatletter
\relax
\def\mn@urlcharsother{\let\do\@makeother \do\$\do\&\do\#\do\^\do\_\do\%\do\~}
\def\mn@doi{\begingroup\mn@urlcharsother \@ifnextchar [ {\mn@doi@} {\mn@doi@[]}}
\def\mn@doi@[#1]#2{\def\@tempa{#1}\ifx\@tempa\@empty \href {http://dx.doi.org/#2} {doi:#2}\else \href {http://dx.doi.org/#2} {#1}\fi \endgroup}
\def\mn@eprint#1#2{\mn@eprint@#1:#2::\@nil}
\def\mn@eprint@arXiv#1{\href {http://arxiv.org/abs/#1} {{\tt arXiv:#1}}}
\def\mn@eprint@dblp#1{\href {http://dblp.uni-trier.de/rec/bibtex/#1.xml} {dblp:#1}}
\def\mn@eprint@#1:#2:#3:#4\@nil{\def\@tempa {#1}\def\@tempb {#2}\def\@tempc {#3}\ifx \@tempc \@empty \let \@tempc \@tempb \let \@tempb \@tempa \fi \ifx \@tempb \@empty \def\@tempb {arXiv}\fi \@ifundefined {mn@eprint@\@tempb}{\@tempb:\@tempc}{\expandafter \expandafter \csname mn@eprint@\@tempb\endcsname \expandafter{\@tempc}}}

\bibitem[\protect\citeauthoryear{{Alibert}}{{Alibert}}{2014}]{Alibert2014}
{Alibert} Y.,  2014, \mn@doi [\aap] {10.1051/0004-6361/201322293}, \href {https://ui.adsabs.harvard.edu/abs/2014A&A...561A..41A} {561, A41}

\bibitem[\protect\citeauthoryear{{Anderson}, {Dubrovinsky}, {Saxena}  \& {LeBihan}}{{Anderson} et~al.}{2001}]{Anderson2001}
{Anderson} O.~L.,  {Dubrovinsky} L.,  {Saxena} S.~K.,   {LeBihan} T.,  2001, \mn@doi [\grl] {10.1029/2000GL008544}, \href {https://ui.adsabs.harvard.edu/abs/2001GeoRL..28..399A} {28, 399}

\bibitem[\protect\citeauthoryear{{Benneke} et~al.,}{{Benneke} et~al.}{2017}]{Benneke2017}
{Benneke} B.,  et~al., 2017, \mn@doi [\apj] {10.3847/1538-4357/834/2/187}, \href {https://ui.adsabs.harvard.edu/abs/2017ApJ...834..187B} {834, 187}

\bibitem[\protect\citeauthoryear{{Benneke} et~al.,}{{Benneke} et~al.}{2019}]{Benneke2019}
{Benneke} B.,  et~al., 2019, \mn@doi [\apjl] {10.3847/2041-8213/ab59dc}, \href {https://ui.adsabs.harvard.edu/abs/2019ApJ...887L..14B} {887, L14}

\bibitem[\protect\citeauthoryear{{Birch}}{{Birch}}{1952}]{Birch1952}
{Birch} F.,  1952, \mn@doi [\jgr] {10.1029/JZ057i002p00227}, \href {https://ui.adsabs.harvard.edu/abs/1952JGR....57..227B} {57, 227}

\bibitem[\protect\citeauthoryear{{Bitsch}, {Raymond}, {Buchhave}, {Bello-Arufe}, {Rathcke}  \& {Schneider}}{{Bitsch} et~al.}{2021}]{Bitsch2021}
{Bitsch} B.,  {Raymond} S.~N.,  {Buchhave} L.~A.,  {Bello-Arufe} A.,  {Rathcke} A.~D.,   {Schneider} A.~D.,  2021, \mn@doi [\aap] {10.1051/0004-6361/202140793}, \href {https://ui.adsabs.harvard.edu/abs/2021A&A...649L...5B} {649, L5}

\bibitem[\protect\citeauthoryear{{Blain}, {Charnay}  \& {B{\'e}zard}}{{Blain} et~al.}{2021}]{Blain2021}
{Blain} D.,  {Charnay} B.,   {B{\'e}zard} B.,  2021, \mn@doi [Astron. Astrophys.] {10.1051/0004-6361/202039072}, \href {https://ui.adsabs.harvard.edu/abs/2021A&A...646A..15B} {646, A15}

\bibitem[\protect\citeauthoryear{{Bonfanti} et~al.,}{{Bonfanti} et~al.}{2024}]{Bonfanti2024}
{Bonfanti} A.,  et~al., 2024, \mn@doi [\aap] {10.1051/0004-6361/202348180}, \href {https://ui.adsabs.harvard.edu/abs/2024A&A...682A..66B} {682, A66}

\bibitem[\protect\citeauthoryear{{Brugger}, {Mousis}, {Deleuil}  \& {Deschamps}}{{Brugger} et~al.}{2017}]{Brugger2017}
{Brugger} B.,  {Mousis} O.,  {Deleuil} M.,   {Deschamps} F.,  2017, \mn@doi [\apj] {10.3847/1538-4357/aa965a}, \href {https://ui.adsabs.harvard.edu/abs/2017ApJ...850...93B} {850, 93}

\bibitem[\protect\citeauthoryear{{Cadieux} et~al.,}{{Cadieux} et~al.}{2024}]{Cadieux2024}
{Cadieux} C.,  et~al., 2024, \mn@doi [\apjl] {10.3847/2041-8213/ad1691}, \href {https://ui.adsabs.harvard.edu/abs/2024ApJ...960L...3C} {960, L3}

\bibitem[\protect\citeauthoryear{{Caillabet}, {Mazevet}  \& {Loubeyre}}{{Caillabet} et~al.}{2011}]{Caillabet2011}
{Caillabet} L.,  {Mazevet} S.,   {Loubeyre} P.,  2011, \mn@doi [\prb] {10.1103/PhysRevB.83.094101}, \href {https://ui.adsabs.harvard.edu/abs/2011PhRvB..83i4101C} {83, 094101}

\bibitem[\protect\citeauthoryear{{Chabrier} \& {Potekhin}}{{Chabrier} \& {Potekhin}}{1998}]{Chabrier1998}
{Chabrier} G.,  {Potekhin} A.~Y.,  1998, \mn@doi [\pre] {10.1103/PhysRevE.58.4941}, \href {https://ui.adsabs.harvard.edu/abs/1998PhRvE..58.4941C} {58, 4941}

\bibitem[\protect\citeauthoryear{{Chabrier}, {Mazevet}  \& {Soubiran}}{{Chabrier} et~al.}{2019}]{Chabrier2019}
{Chabrier} G.,  {Mazevet} S.,   {Soubiran} F.,  2019, \mn@doi [\apj] {10.3847/1538-4357/aaf99f}, \href {https://ui.adsabs.harvard.edu/abs/2019ApJ...872...51C} {872, 51}

\bibitem[\protect\citeauthoryear{Charette \& Smith}{Charette \& Smith}{2010}]{Charette2010}
Charette M.,  Smith W.,  2010, Oceanography, 23, 112–114

\bibitem[\protect\citeauthoryear{Chaturvedi et~al.,}{Chaturvedi et~al.}{2022}]{Chaturvedi2022}
Chaturvedi P.,  et~al., 2022, \mn@doi [Astronomy \& Astrophysics] {10.1051/0004-6361/202244056}, 666, A155

\bibitem[\protect\citeauthoryear{{Choblet}, {Tobie}, {Sotin}, {Kalousov{\'a}}  \& {Grasset}}{{Choblet} et~al.}{2017}]{Choblet2017}
{Choblet} G.,  {Tobie} G.,  {Sotin} C.,  {Kalousov{\'a}} K.,   {Grasset} O.,  2017, \mn@doi [\icarus] {10.1016/j.icarus.2016.12.002}, \href {https://ui.adsabs.harvard.edu/abs/2017Icar..285..252C} {285, 252}

\bibitem[\protect\citeauthoryear{{Cloutier} \& {Menou}}{{Cloutier} \& {Menou}}{2020}]{Cloutier2020b}
{Cloutier} R.,  {Menou} K.,  2020, \mn@doi [\aj] {10.3847/1538-3881/ab8237}, \href {https://ui.adsabs.harvard.edu/abs/2020AJ....159..211C} {159, 211}

\bibitem[\protect\citeauthoryear{{Cloutier} et~al.,}{{Cloutier} et~al.}{2019}]{Cloutier2019}
{Cloutier} R.,  et~al., 2019, \mn@doi [\aap] {10.1051/0004-6361/201833995}, \href {https://ui.adsabs.harvard.edu/abs/2019A&A...621A..49C} {621, A49}

\bibitem[\protect\citeauthoryear{{Cloutier} et~al.,}{{Cloutier} et~al.}{2020}]{Cloutier2020}
{Cloutier} R.,  et~al., 2020, \mn@doi [\aj] {10.3847/1538-3881/ab91c2}, \href {https://ui.adsabs.harvard.edu/abs/2020AJ....160....3C} {160, 3}

\bibitem[\protect\citeauthoryear{{Dittmann} et~al.,}{{Dittmann} et~al.}{2017}]{Dittmann2017}
{Dittmann} J.~A.,  et~al., 2017, \mn@doi [\nat] {10.1038/nature22055}, \href {https://ui.adsabs.harvard.edu/abs/2017Natur.544..333D} {544, 333}

\bibitem[\protect\citeauthoryear{{Dorn} \& {Lichtenberg}}{{Dorn} \& {Lichtenberg}}{2021}]{Dorn2021}
{Dorn} C.,  {Lichtenberg} T.,  2021, \mn@doi [\apjl] {10.3847/2041-8213/ac33af}, \href {https://ui.adsabs.harvard.edu/abs/2021ApJ...922L...4D} {922, L4}

\bibitem[\protect\citeauthoryear{{Dorn}, {Venturini}, {Khan}, {Heng}, {Alibert}, {Helled}, {Rivoldini}  \& {Benz}}{{Dorn} et~al.}{2017}]{Dorn2017}
{Dorn} C.,  {Venturini} J.,  {Khan} A.,  {Heng} K.,  {Alibert} Y.,  {Helled} R.,  {Rivoldini} A.,   {Benz} W.,  2017, \mn@doi [\aap] {10.1051/0004-6361/201628708}, \href {https://ui.adsabs.harvard.edu/abs/2017A&A...597A..37D} {597, A37}

\bibitem[\protect\citeauthoryear{{Dunaeva}, {Antsyshkin}  \& {Kuskov}}{{Dunaeva} et~al.}{2010}]{Dunaeva2010}
{Dunaeva} A.~N.,  {Antsyshkin} D.~V.,   {Kuskov} O.~L.,  2010, \mn@doi [Solar System Research] {10.1134/S0038094610030044}, \href {https://ui.adsabs.harvard.edu/abs/2010SoSyR..44..202D} {44, 202}

\bibitem[\protect\citeauthoryear{{Elkins-Tanton} \& {Seager}}{{Elkins-Tanton} \& {Seager}}{2008}]{ElkinsTanton2008}
{Elkins-Tanton} L.~T.,  {Seager} S.,  2008, \mn@doi [\apj] {10.1086/591433}, \href {https://ui.adsabs.harvard.edu/abs/2008ApJ...685.1237E} {685, 1237}

\bibitem[\protect\citeauthoryear{Fei, Mao  \& Hemley}{Fei et~al.}{1993}]{Fei1993}
Fei Y.,  Mao H.,   Hemley R.~J.,  1993, J. Chem. Phys., 99, 5369

\bibitem[\protect\citeauthoryear{Feistel \& Wagner}{Feistel \& Wagner}{2006}]{Feistel2006}
Feistel R.,  Wagner W.,  2006, Journal of Physical and Chemical Reference Data, 35, 1021

\bibitem[\protect\citeauthoryear{{Fortney}, {Marley}  \& {Barnes}}{{Fortney} et~al.}{2007}]{Fortney2007}
{Fortney} J.~J.,  {Marley} M.~S.,   {Barnes} J.~W.,  2007, \mn@doi [\apj] {10.1086/512120}, \href {https://ui.adsabs.harvard.edu/abs/2007ApJ...659.1661F} {659, 1661}

\bibitem[\protect\citeauthoryear{{French}, {Mattsson}, {Nettelmann}  \& {Redmer}}{{French} et~al.}{2009}]{French2009}
{French} M.,  {Mattsson} T.~R.,  {Nettelmann} N.,   {Redmer} R.,  2009, \mn@doi [\prb] {10.1103/PhysRevB.79.054107}, \href {https://ui.adsabs.harvard.edu/abs/2009PhRvB..79e4107F} {79, 054107}

\bibitem[\protect\citeauthoryear{{Fulton} \& {Petigura}}{{Fulton} \& {Petigura}}{2018}]{Fulton2018}
{Fulton} B.~J.,  {Petigura} E.~A.,  2018, \mn@doi [\aj] {10.3847/1538-3881/aae828}, \href {https://ui.adsabs.harvard.edu/abs/2018AJ....156..264F} {156, 264}

\bibitem[\protect\citeauthoryear{{Fulton} et~al.,}{{Fulton} et~al.}{2017}]{Fulton2017}
{Fulton} B.~J.,  et~al., 2017, \mn@doi [\aj] {10.3847/1538-3881/aa80eb}, \href {https://ui.adsabs.harvard.edu/abs/2017AJ....154..109F} {154, 109}

\bibitem[\protect\citeauthoryear{{Gandhi} \& {Madhusudhan}}{{Gandhi} \& {Madhusudhan}}{2017}]{Gandhi2017}
{Gandhi} S.,  {Madhusudhan} N.,  2017, \mn@doi [\mnras] {10.1093/mnras/stx1601}, \href {https://ui.adsabs.harvard.edu/abs/2017MNRAS.472.2334G} {472, 2334}

\bibitem[\protect\citeauthoryear{Gardner, Armstrong, Calder  \& Beaudoin}{Gardner et~al.}{2014}]{Gardner2014}
Gardner J.~V.,  Armstrong A.~A.,  Calder B.~R.,   Beaudoin J.,  2014, \mn@doi [Marine Geodesy] {10.1080/01490419.2013.837849}, 37, 1

\bibitem[\protect\citeauthoryear{{Genda}}{{Genda}}{2016}]{Hidenori2016}
{Genda} H.,  2016, \mn@doi [Geochemical Journal] {10.2343/geochemj.2.0398}, \href {https://ui.adsabs.harvard.edu/abs/2016GeocJ..50...27G} {50, 27}

\bibitem[\protect\citeauthoryear{{Grande} et~al.,}{{Grande} et~al.}{2022}]{Grande2022}
{Grande} Z.~M.,  et~al., 2022, \mn@doi [\prb] {10.1103/PhysRevB.105.104109}, \href {https://ui.adsabs.harvard.edu/abs/2022PhRvB.105j4109G} {105, 104109}

\bibitem[\protect\citeauthoryear{{Grasset}, {Schneider}  \& {Sotin}}{{Grasset} et~al.}{2009}]{Grasset2009}
{Grasset} O.,  {Schneider} J.,   {Sotin} C.,  2009, \mn@doi [\apj] {10.1088/0004-637X/693/1/722}, \href {https://ui.adsabs.harvard.edu/abs/2009ApJ...693..722G} {693, 722}

\bibitem[\protect\citeauthoryear{{G{\"u}nther} et~al.,}{{G{\"u}nther} et~al.}{2019}]{Gunther2019}
{G{\"u}nther} M.~N.,  et~al., 2019, \mn@doi [Nature Astronomy] {10.1038/s41550-019-0845-5}, \href {https://ui.adsabs.harvard.edu/abs/2019NatAs...3.1099G} {3, 1099}

\bibitem[\protect\citeauthoryear{{Gupta} \& {Schlichting}}{{Gupta} \& {Schlichting}}{2019}]{Gupta2019}
{Gupta} A.,  {Schlichting} H.~E.,  2019, \mn@doi [\mnras] {10.1093/mnras/stz1230}, \href {https://ui.adsabs.harvard.edu/abs/2019MNRAS.487...24G} {487, 24}

\bibitem[\protect\citeauthoryear{{Gupta} \& {Schlichting}}{{Gupta} \& {Schlichting}}{2020}]{Gupta2020}
{Gupta} A.,  {Schlichting} H.~E.,  2020, \mn@doi [\mnras] {10.1093/mnras/staa315}, \href {https://ui.adsabs.harvard.edu/abs/2020MNRAS.493..792G} {493, 792}

\bibitem[\protect\citeauthoryear{{Haldemann}, {Alibert}, {Mordasini}  \& {Benz}}{{Haldemann} et~al.}{2020}]{Haldemann2020}
{Haldemann} J.,  {Alibert} Y.,  {Mordasini} C.,   {Benz} W.,  2020, \mn@doi [\aap] {10.1051/0004-6361/202038367}, \href {https://ui.adsabs.harvard.edu/abs/2020A&A...643A.105H} {643, A105}

\bibitem[\protect\citeauthoryear{{Hardegree-Ullman}, {Zink}, {Christiansen}, {Dressing}, {Ciardi}  \& {Schlieder}}{{Hardegree-Ullman} et~al.}{2020}]{Hardegree2020}
{Hardegree-Ullman} K.~K.,  {Zink} J.~K.,  {Christiansen} J.~L.,  {Dressing} C.~D.,  {Ciardi} D.~R.,   {Schlieder} J.~E.,  2020, \mn@doi [\apjs] {10.3847/1538-4365/ab7230}, \href {https://ui.adsabs.harvard.edu/abs/2020ApJS..247...28H} {247, 28}

\bibitem[\protect\citeauthoryear{Hemley et~al.}{Hemley et~al.}{1987}]{Hemley1987}
Hemley R.,  et~al., 1987, Nature, 330, 737–740

\bibitem[\protect\citeauthoryear{{Hernandez}, {Caracas}  \& {Labrosse}}{{Hernandez} et~al.}{2022}]{Hernandez2022}
{Hernandez} J.-A.,  {Caracas} R.,   {Labrosse} S.,  2022, \mn@doi [Nature Communications] {10.1038/s41467-022-30796-5}, \href {https://ui.adsabs.harvard.edu/abs/2022NatCo..13.3303H} {13, 3303}

\bibitem[\protect\citeauthoryear{{Hu}, {Damiano}, {Scheucher}, {Kite}, {Seager}  \& {Rauer}}{{Hu} et~al.}{2021}]{Hu2021}
{Hu} R.,  {Damiano} M.,  {Scheucher} M.,  {Kite} E.,  {Seager} S.,   {Rauer} H.,  2021, \mn@doi [\apjl] {10.3847/2041-8213/ac1f92}, \href {https://ui.adsabs.harvard.edu/abs/2021ApJ...921L...8H} {921, L8}

\bibitem[\protect\citeauthoryear{{Huang} et~al.,}{{Huang} et~al.}{2021}]{Huang2021}
{Huang} C.,  et~al., 2021, \mn@doi [\mnras] {10.1093/mnras/stab645}, \href {https://ui.adsabs.harvard.edu/abs/2021MNRAS.503.2825H} {503, 2825}

\bibitem[\protect\citeauthoryear{{Huang}, {Rice}  \& {Steffen}}{{Huang} et~al.}{2022}]{Huang2022}
{Huang} C.,  {Rice} D.~R.,   {Steffen} J.~H.,  2022, \mn@doi [\mnras] {10.1093/mnras/stac1133}, \href {https://ui.adsabs.harvard.edu/abs/2022MNRAS.513.5256H} {513, 5256}

\bibitem[\protect\citeauthoryear{{Izidoro}, {Schlichting}, {Isella}, {Dasgupta}, {Zimmermann}  \& {Bitsch}}{{Izidoro} et~al.}{2022}]{Izidoro2022}
{Izidoro} A.,  {Schlichting} H.~E.,  {Isella} A.,  {Dasgupta} R.,  {Zimmermann} C.,   {Bitsch} B.,  2022, \mn@doi [\apjl] {10.3847/2041-8213/ac990d}, \href {https://ui.adsabs.harvard.edu/abs/2022ApJ...939L..19I} {939, L19}

\bibitem[\protect\citeauthoryear{{Journaux} et~al.,}{{Journaux} et~al.}{2020a}]{Journaux2020a}
{Journaux} B.,  et~al., 2020a, \mn@doi [Journal of Geophysical Research (Planets)] {10.1029/2019JE006176}, \href {https://ui.adsabs.harvard.edu/abs/2020JGRE..12506176J} {125, e06176}

\bibitem[\protect\citeauthoryear{{Journaux} et~al.,}{{Journaux} et~al.}{2020b}]{Journaux2020b}
{Journaux} B.,  et~al., 2020b, \mn@doi [\ssr] {10.1007/s11214-019-0633-7}, \href {https://ui.adsabs.harvard.edu/abs/2020SSRv..216....7J} {216, 7}

\bibitem[\protect\citeauthoryear{Kalousová \& Sotin}{Kalousová \& Sotin}{2018}]{Kalousova2018b}
Kalousová K.,  Sotin C.,  2018, \mn@doi [Geophys. Res. Lett.] {10.1029/2018GL078889}, 45, 8096

\bibitem[\protect\citeauthoryear{Kalousová et~al.}{Kalousová et~al.}{2018}]{Kalousova2018a}
Kalousová K.,  et~al., 2018, \mn@doi [Icarus] {10.1016/J.ICARUS.2017.07.018}, 299, 133

\bibitem[\protect\citeauthoryear{Karki et~al.}{Karki et~al.}{2000}]{Karki2000}
Karki B.~B.,  et~al., 2000, Phys. Rev. B, 62, 14750

\bibitem[\protect\citeauthoryear{{Kasting}, {Whitmire}  \& {Reynolds}}{{Kasting} et~al.}{1993}]{Kasting1993}
{Kasting} J.~F.,  {Whitmire} D.~P.,   {Reynolds} R.~T.,  1993, \mn@doi [\icarus] {10.1006/icar.1993.1010}, \href {https://ui.adsabs.harvard.edu/abs/1993Icar..101..108K} {101, 108}

\bibitem[\protect\citeauthoryear{{Kaye} et~al.,}{{Kaye} et~al.}{2022}]{Kaye2022}
{Kaye} L.,  et~al., 2022, \mn@doi [\mnras] {10.1093/mnras/stab3483}, \href {https://ui.adsabs.harvard.edu/abs/2022MNRAS.510.5464K} {510, 5464}

\bibitem[\protect\citeauthoryear{{Kite} \& {Ford}}{{Kite} \& {Ford}}{2018}]{Kite2018}
{Kite} E.~S.,  {Ford} E.~B.,  2018, \mn@doi [\apj] {10.3847/1538-4357/aad6e0}, \href {https://ui.adsabs.harvard.edu/abs/2018ApJ...864...75K} {864, 75}

\bibitem[\protect\citeauthoryear{{Kitzmann} et~al.,}{{Kitzmann} et~al.}{2015}]{Kitzmann2015}
{Kitzmann} D.,  et~al., 2015, \mn@doi [\mnras] {10.1093/mnras/stv1487}, \href {https://ui.adsabs.harvard.edu/abs/2015MNRAS.452.3752K} {452, 3752}

\bibitem[\protect\citeauthoryear{Klotz et~al.}{Klotz et~al.}{2017}]{Klotz2017}
Klotz S.,  et~al., 2017, Phys. Rev. B, 95, 174111

\bibitem[\protect\citeauthoryear{Knudson et~al.}{Knudson et~al.}{2012}]{Knudson2012}
Knudson M.~D.,  et~al., 2012, Phys. Rev. Lett., 108, 091102

\bibitem[\protect\citeauthoryear{{Kova{\v{c}}evi{\'c}}, {Gonz{\'a}lez-Cataldo}, {Stewart}  \& {Militzer}}{{Kova{\v{c}}evi{\'c}} et~al.}{2022}]{Kovacevic2022}
{Kova{\v{c}}evi{\'c}} T.,  {Gonz{\'a}lez-Cataldo} F.,  {Stewart} S.~T.,   {Militzer} B.,  2022, \mn@doi [Scientific Reports] {10.1038/s41598-022-16816-w}, \href {https://ui.adsabs.harvard.edu/abs/2022NatSR..1213055K} {12, 13055}

\bibitem[\protect\citeauthoryear{{Lebec} et~al.}{{Lebec} et~al.}{2023}]{Lebec2023}
{Lebec} L.,  et~al., 2023, \mn@doi [Icarus] {10.1016/j.icarus.2023.115494}, \href {https://ui.adsabs.harvard.edu/abs/2023Icar..39615494L} {396, 115494}

\bibitem[\protect\citeauthoryear{{L{\'e}ger} et~al.,}{{L{\'e}ger} et~al.}{2004}]{Leger2004}
{L{\'e}ger} A.,  et~al., 2004, \mn@doi [\icarus] {10.1016/j.icarus.2004.01.001}, \href {https://ui.adsabs.harvard.edu/abs/2004Icar..169..499L} {169, 499}

\bibitem[\protect\citeauthoryear{{Leleu} et~al.,}{{Leleu} et~al.}{2021}]{Leleu2021}
{Leleu} A.,  et~al., 2021, \mn@doi [\aap] {10.1051/0004-6361/202039767}, \href {https://ui.adsabs.harvard.edu/abs/2021A&A...649A..26L} {649, A26}

\bibitem[\protect\citeauthoryear{{Levi}, {Sasselov}  \& {Podolak}}{{Levi} et~al.}{2017}]{Levi2017}
{Levi} A.,  {Sasselov} D.,   {Podolak} M.,  2017, \mn@doi [\apj] {10.3847/1538-4357/aa5cfe}, \href {https://ui.adsabs.harvard.edu/abs/2017ApJ...838...24L} {838, 24}

\bibitem[\protect\citeauthoryear{{Lillo-Box} et~al.,}{{Lillo-Box} et~al.}{2020}]{Lillo2020}
{Lillo-Box} J.,  et~al., 2020, \mn@doi [\aap] {10.1051/0004-6361/202038922}, \href {https://ui.adsabs.harvard.edu/abs/2020A&A...642A.121L} {642, A121}

\bibitem[\protect\citeauthoryear{{Lingam} \& {Loeb}}{{Lingam} \& {Loeb}}{2018}]{Lingam2018}
{Lingam} M.,  {Loeb} A.,  2018, \mn@doi [Astron. J.] {10.3847/1538-3881/aada02}, \href {https://ui.adsabs.harvard.edu/abs/2018AJ....156..151L} {156, 151}

\bibitem[\protect\citeauthoryear{{Lopez}, {Fortney}  \& {Miller}}{{Lopez} et~al.}{2012}]{Lopez2012}
{Lopez} E.~D.,  {Fortney} J.~J.,   {Miller} N.,  2012, \mn@doi [\apj] {10.1088/0004-637X/761/1/59}, \href {https://ui.adsabs.harvard.edu/abs/2012ApJ...761...59L} {761, 59}

\bibitem[\protect\citeauthoryear{{Madhusudhan}, {Lee}  \& {Mousis}}{{Madhusudhan} et~al.}{2012}]{Madhusudhan2012}
{Madhusudhan} N.,  {Lee} K. K.~M.,   {Mousis} O.,  2012, \mn@doi [\apjl] {10.1088/2041-8205/759/2/L40}, \href {https://ui.adsabs.harvard.edu/abs/2012ApJ...759L..40M} {759, L40}

\bibitem[\protect\citeauthoryear{{Madhusudhan}, {Nixon}, {Welbanks}, {Piette}  \& {Booth}}{{Madhusudhan} et~al.}{2020}]{Madhusudhan2020}
{Madhusudhan} N.,  {Nixon} M.~C.,  {Welbanks} L.,  {Piette} A. A.~A.,   {Booth} R.~A.,  2020, \mn@doi [\apjl] {10.3847/2041-8213/ab7229}, \href {https://ui.adsabs.harvard.edu/abs/2020ApJ...891L...7M} {891, L7}

\bibitem[\protect\citeauthoryear{{Madhusudhan}, {Piette}  \& {Constantinou}}{{Madhusudhan} et~al.}{2021}]{Madhusudhan2021}
{Madhusudhan} N.,  {Piette} A. A.~A.,   {Constantinou} S.,  2021, \mn@doi [\apj] {10.3847/1538-4357/abfd9c}, \href {https://ui.adsabs.harvard.edu/abs/2021ApJ...918....1M} {918, 1}

\bibitem[\protect\citeauthoryear{{Madhusudhan}, {Moses}, {Rigby}  \& {Barrier}}{{Madhusudhan} et~al.}{2023a}]{Madhusudhan2023a}
{Madhusudhan} N.,  {Moses} J.~I.,  {Rigby} F.,   {Barrier} E.,  2023a, \mn@doi [Faraday Discussions] {10.1039/D3FD00075C}, \href {https://ui.adsabs.harvard.edu/abs/2023FaDi..245...80M} {245, 80}

\bibitem[\protect\citeauthoryear{{Madhusudhan}, {Sarkar}, {Constantinou}, {Holmberg}, {Piette}  \& {Moses}}{{Madhusudhan} et~al.}{2023b}]{Madhusudhan2023b}
{Madhusudhan} N.,  {Sarkar} S.,  {Constantinou} S.,  {Holmberg} M.,  {Piette} A. A.~A.,   {Moses} J.~I.,  2023b, \mn@doi [\apjl] {10.3847/2041-8213/acf577}, \href {https://ui.adsabs.harvard.edu/abs/2023ApJ...956L..13M} {956, L13}

\bibitem[\protect\citeauthoryear{Maruyama et~al.}{Maruyama et~al.}{2013}]{Maruyama2013}
Maruyama S.,  et~al., 2013, \mn@doi [Geosci. Front.] {10.1016/J.GSF.2012.11.001}, 4, 141

\bibitem[\protect\citeauthoryear{{Meadows} \& {Barnes}}{{Meadows} \& {Barnes}}{2018}]{Meadows2018}
{Meadows} V.~S.,  {Barnes} R.~K.,  2018, in {Deeg} H.~J.,  {Belmonte} J.~A.,  eds, , Handbook of Exoplanets.
p.~57, \mn@doi{10.1007/978-3-319-55333-7_57}

\bibitem[\protect\citeauthoryear{{Ment} et~al.,}{{Ment} et~al.}{2019}]{Ment2019}
{Ment} K.,  et~al., 2019, \mn@doi [\aj] {10.3847/1538-3881/aaf1b1}, \href {https://ui.adsabs.harvard.edu/abs/2019AJ....157...32M} {157, 32}

\bibitem[\protect\citeauthoryear{Merino et~al.}{Merino et~al.}{2019}]{Merino2019}
Merino N.,  et~al., 2019, Front. Microbiol., 10, 780

\bibitem[\protect\citeauthoryear{{Mikal-Evans} et~al.,}{{Mikal-Evans} et~al.}{2023}]{MikalEvans2023}
{Mikal-Evans} T.,  et~al., 2023, \mn@doi [\aj] {10.3847/1538-3881/aca90b}, \href {https://ui.adsabs.harvard.edu/abs/2023AJ....165...84M} {165, 84}

\bibitem[\protect\citeauthoryear{{Mousis}, {Deleuil}, {Aguichine}, {Marcq}, {Naar}, {Aguirre}, {Brugger}  \& {Gon{\c{c}}alves}}{{Mousis} et~al.}{2020}]{Mousis2020}
{Mousis} O.,  {Deleuil} M.,  {Aguichine} A.,  {Marcq} E.,  {Naar} J.,  {Aguirre} L.~A.,  {Brugger} B.,   {Gon{\c{c}}alves} T.,  2020, \mn@doi [\apjl] {10.3847/2041-8213/ab9530}, \href {https://ui.adsabs.harvard.edu/abs/2020ApJ...896L..22M} {896, L22}

\bibitem[\protect\citeauthoryear{{Nixon} \& {Madhusudhan}}{{Nixon} \& {Madhusudhan}}{2021}]{Nixon2021}
{Nixon} M.~C.,  {Madhusudhan} N.,  2021, \mn@doi [\mnras] {10.1093/mnras/stab1500}, \href {https://ui.adsabs.harvard.edu/abs/2021MNRAS.505.3414N} {505, 3414}

\bibitem[\protect\citeauthoryear{{Noack} et~al.,}{{Noack} et~al.}{2016}]{Noack2016}
{Noack} L.,  et~al., 2016, \mn@doi [\icarus] {10.1016/j.icarus.2016.05.009}, \href {https://ui.adsabs.harvard.edu/abs/2016Icar..277..215N} {277, 215}

\bibitem[\protect\citeauthoryear{{Nowak} et~al.,}{{Nowak} et~al.}{2020}]{Nowak2020}
{Nowak} G.,  et~al., 2020, \mn@doi [\aap] {10.1051/0004-6361/202037867}, \href {https://ui.adsabs.harvard.edu/abs/2020A&A...642A.173N} {642, A173}

\bibitem[\protect\citeauthoryear{{O'Malley-James} \& {Kaltenegger}}{{O'Malley-James} \& {Kaltenegger}}{2017}]{OMalleyJames2017}
{O'Malley-James} J.~T.,  {Kaltenegger} L.,  2017, \mn@doi [\mnras] {10.1093/mnrasl/slx047}, \href {https://ui.adsabs.harvard.edu/abs/2017MNRAS.469L..26O} {469, L26}

\bibitem[\protect\citeauthoryear{{Owen} \& {Wu}}{{Owen} \& {Wu}}{2013}]{Owen2013}
{Owen} J.~E.,  {Wu} Y.,  2013, \mn@doi [\apj] {10.1088/0004-637X/775/2/105}, \href {https://ui.adsabs.harvard.edu/abs/2013ApJ...775..105O} {775, 105}

\bibitem[\protect\citeauthoryear{{Owen} \& {Wu}}{{Owen} \& {Wu}}{2017}]{Owen2017}
{Owen} J.~E.,  {Wu} Y.,  2017, \mn@doi [\apj] {10.3847/1538-4357/aa890a}, \href {https://ui.adsabs.harvard.edu/abs/2017ApJ...847...29O} {847, 29}

\bibitem[\protect\citeauthoryear{{Petigura}}{{Petigura}}{2020}]{Petigura2020}
{Petigura} E.~A.,  2020, \mn@doi [\aj] {10.3847/1538-3881/ab9fff}, \href {https://ui.adsabs.harvard.edu/abs/2020AJ....160...89P} {160, 89}

\bibitem[\protect\citeauthoryear{{Piette} \& {Madhusudhan}}{{Piette} \& {Madhusudhan}}{2020}]{Piette2020}
{Piette} A. A.~A.,  {Madhusudhan} N.,  2020, \mn@doi [\apj] {10.3847/1538-4357/abbfb1}, \href {https://ui.adsabs.harvard.edu/abs/2020ApJ...904..154P} {904, 154}

\bibitem[\protect\citeauthoryear{{Podolak}, {Helled}  \& {Schubert}}{{Podolak} et~al.}{2019}]{Podolak2019}
{Podolak} M.,  {Helled} R.,   {Schubert} G.,  2019, \mn@doi [\mnras] {10.1093/mnras/stz1467}, \href {https://ui.adsabs.harvard.edu/abs/2019MNRAS.487.2653P} {487, 2653}

\bibitem[\protect\citeauthoryear{{Ramirez}}{{Ramirez}}{2018}]{Ramirez2018b}
{Ramirez} R.~M.,  2018, \mn@doi [Geosciences] {10.3390/geosciences8080280}, \href {https://ui.adsabs.harvard.edu/abs/2018Geosc...8..280R} {8, 280}

\bibitem[\protect\citeauthoryear{{Ramirez} \& {Levi}}{{Ramirez} \& {Levi}}{2018}]{Ramirez2018}
{Ramirez} R.~M.,  {Levi} A.,  2018, \mn@doi [\mnras] {10.1093/mnras/sty761}, \href {https://ui.adsabs.harvard.edu/abs/2018MNRAS.477.4627R} {477, 4627}

\bibitem[\protect\citeauthoryear{{Ricker} et~al.,}{{Ricker} et~al.}{2015}]{Ricker2015}
{Ricker} G.~R.,  et~al., 2015, \mn@doi [Journal of Astronomical Telescopes, Instruments, and Systems] {10.1117/1.JATIS.1.1.014003}, \href {https://ui.adsabs.harvard.edu/abs/2015JATIS...1a4003R} {1, 014003}

\bibitem[\protect\citeauthoryear{{Rogers} \& {Owen}}{{Rogers} \& {Owen}}{2021}]{Rogers2021}
{Rogers} J.~G.,  {Owen} J.~E.,  2021, \mn@doi [\mnras] {10.1093/mnras/stab529}, \href {https://ui.adsabs.harvard.edu/abs/2021MNRAS.503.1526R} {503, 1526}

\bibitem[\protect\citeauthoryear{{Rogers} \& {Seager}}{{Rogers} \& {Seager}}{2010a}]{Rogers2010a}
{Rogers} L.~A.,  {Seager} S.,  2010a, \mn@doi [\apj] {10.1088/0004-637X/712/2/974}, \href {https://ui.adsabs.harvard.edu/abs/2010ApJ...712..974R} {712, 974}

\bibitem[\protect\citeauthoryear{{Rogers} \& {Seager}}{{Rogers} \& {Seager}}{2010b}]{Rogers2010b}
{Rogers} L.~A.,  {Seager} S.,  2010b, \mn@doi [\apj] {10.1088/0004-637X/716/2/1208}, \href {https://ui.adsabs.harvard.edu/abs/2010ApJ...716.1208R} {716, 1208}

\bibitem[\protect\citeauthoryear{{Rogers}, {Bodenheimer}, {Lissauer}  \& {Seager}}{{Rogers} et~al.}{2011}]{Rogers2011}
{Rogers} L.~A.,  {Bodenheimer} P.,  {Lissauer} J.~J.,   {Seager} S.,  2011, \mn@doi [\apj] {10.1088/0004-637X/738/1/59}, \href {https://ui.adsabs.harvard.edu/abs/2011ApJ...738...59R} {738, 59}

\bibitem[\protect\citeauthoryear{{Rogers}, {Schlichting}  \& {Owen}}{{Rogers} et~al.}{2023}]{Rogers2023}
{Rogers} J.~G.,  {Schlichting} H.~E.,   {Owen} J.~E.,  2023, \mn@doi [\apjl] {10.3847/2041-8213/acc86f}, \href {https://ui.adsabs.harvard.edu/abs/2023ApJ...947L..19R} {947, L19}

\bibitem[\protect\citeauthoryear{Rothschild \& Mancinelli}{Rothschild \& Mancinelli}{2001}]{Rothschild2001}
Rothschild L.~J.,  Mancinelli R.~L.,  2001, Nature, 409, 1092

\bibitem[\protect\citeauthoryear{{Salpeter} \& {Zapolsky}}{{Salpeter} \& {Zapolsky}}{1967}]{Salpeter1967}
{Salpeter} E.~E.,  {Zapolsky} H.~S.,  1967, \mn@doi [Physical Review] {10.1103/PhysRev.158.876}, \href {https://ui.adsabs.harvard.edu/abs/1967PhRv..158..876S} {158, 876}

\bibitem[\protect\citeauthoryear{{Saumon}, {Chabrier}  \& {van Horn}}{{Saumon} et~al.}{1995}]{Saumon1995}
{Saumon} D.,  {Chabrier} G.,   {van Horn} H.~M.,  1995, \mn@doi [\apjs] {10.1086/192204}, \href {https://ui.adsabs.harvard.edu/abs/1995ApJS...99..713S} {99, 713}

\bibitem[\protect\citeauthoryear{{Seager}, {Kuchner}, {Hier-Majumder}  \& {Militzer}}{{Seager} et~al.}{2007}]{Seager2007}
{Seager} S.,  {Kuchner} M.,  {Hier-Majumder} C.~A.,   {Militzer} B.,  2007, \mn@doi [\apj] {10.1086/521346}, \href {https://ui.adsabs.harvard.edu/abs/2007ApJ...669.1279S} {669, 1279}

\bibitem[\protect\citeauthoryear{{Selsis}, {Kasting}, {Levrard}, {Paillet}, {Ribas}  \& {Delfosse}}{{Selsis} et~al.}{2007}]{Selsis2007}
{Selsis} F.,  {Kasting} J.~F.,  {Levrard} B.,  {Paillet} J.,  {Ribas} I.,   {Delfosse} X.,  2007, \mn@doi [\aap] {10.1051/0004-6361:20078091}, \href {https://ui.adsabs.harvard.edu/abs/2007A&A...476.1373S} {476, 1373}

\bibitem[\protect\citeauthoryear{{Shah}, {Alibert}, {Helled}  \& {Mezger}}{{Shah} et~al.}{2021}]{Shah2021}
{Shah} O.,  {Alibert} Y.,  {Helled} R.,   {Mezger} K.,  2021, \mn@doi [\aap] {10.1051/0004-6361/202038839}, \href {https://ui.adsabs.harvard.edu/abs/2021A&A...646A.162S} {646, A162}

\bibitem[\protect\citeauthoryear{{Shields}, {Ballard}  \& {Johnson}}{{Shields} et~al.}{2016}]{Shields2016}
{Shields} A.~L.,  {Ballard} S.,   {Johnson} J.~A.,  2016, \mn@doi [\physrep] {10.1016/j.physrep.2016.10.003}, \href {https://ui.adsabs.harvard.edu/abs/2016PhR...663....1S} {663, 1}

\bibitem[\protect\citeauthoryear{Sotin, Grasset  \& Mocquet}{Sotin et~al.}{2007}]{Sotin2007}
Sotin C.,  Grasset O.,   Mocquet A.,  2007, Icarus, 191, 337

\bibitem[\protect\citeauthoryear{{Thomas} \& {Madhusudhan}}{{Thomas} \& {Madhusudhan}}{2016}]{Thomas2016a}
{Thomas} S.~W.,  {Madhusudhan} N.,  2016, \mn@doi [\mnras] {10.1093/mnras/stw321}, \href {https://ui.adsabs.harvard.edu/abs/2016MNRAS.458.1330T} {458, 1330}

\bibitem[\protect\citeauthoryear{{Tremblay}, {Line}, {Stevenson}, {Kataria}, {Zellem}, {Fortney}  \& {Morley}}{{Tremblay} et~al.}{2020}]{Tremblay2020}
{Tremblay} L.,  {Line} M.~R.,  {Stevenson} K.,  {Kataria} T.,  {Zellem} R.~T.,  {Fortney} J.~J.,   {Morley} C.,  2020, \mn@doi [\aj] {10.3847/1538-3881/ab64dd}, \href {https://ui.adsabs.harvard.edu/abs/2020AJ....159..117T} {159, 117}

\bibitem[\protect\citeauthoryear{{Tsai}, {Innes}, {Lichtenberg}, {Taylor}, {Malik}, {Chubb}  \& {Pierrehumbert}}{{Tsai} et~al.}{2021}]{Tsai2021}
{Tsai} S.-M.,  {Innes} H.,  {Lichtenberg} T.,  {Taylor} J.,  {Malik} M.,  {Chubb} K.,   {Pierrehumbert} R.,  2021, \mn@doi [\apjl] {10.3847/2041-8213/ac399a}, \href {https://ui.adsabs.harvard.edu/abs/2021ApJ...922L..27T} {922, L27}

\bibitem[\protect\citeauthoryear{{Tsiaras}, {Waldmann}, {Tinetti}, {Tennyson}  \& {Yurchenko}}{{Tsiaras} et~al.}{2019}]{Tsiaras2019}
{Tsiaras} A.,  {Waldmann} I.~P.,  {Tinetti} G.,  {Tennyson} J.,   {Yurchenko} S.~N.,  2019, \mn@doi [Nature Astronomy] {10.1038/s41550-019-0878-9}, \href {https://ui.adsabs.harvard.edu/abs/2019NatAs...3.1086T} {3, 1086}

\bibitem[\protect\citeauthoryear{{Valencia}, {Sasselov}  \& {O'Connell}}{{Valencia} et~al.}{2007}]{Valencia2007}
{Valencia} D.,  {Sasselov} D.~D.,   {O'Connell} R.~J.,  2007, \mn@doi [\apj] {10.1086/519554}, \href {https://ui.adsabs.harvard.edu/abs/2007ApJ...665.1413V} {665, 1413}

\bibitem[\protect\citeauthoryear{{Valencia}, {Guillot}, {Parmentier}  \& {Freedman}}{{Valencia} et~al.}{2013}]{Valencia2013}
{Valencia} D.,  {Guillot} T.,  {Parmentier} V.,   {Freedman} R.~S.,  2013, \mn@doi [\apj] {10.1088/0004-637X/775/1/10}, \href {https://ui.adsabs.harvard.edu/abs/2013ApJ...775...10V} {775, 10}

\bibitem[\protect\citeauthoryear{{Van Eylen} et~al.,}{{Van Eylen} et~al.}{2021}]{VanEylen2021}
{Van Eylen} V.,  et~al., 2021, \mn@doi [\mnras] {10.1093/mnras/stab2143}, \href {https://ui.adsabs.harvard.edu/abs/2021MNRAS.507.2154V} {507, 2154}

\bibitem[\protect\citeauthoryear{{Vazan}, {Sari}  \& {Kessel}}{{Vazan} et~al.}{2022}]{Vazan2022}
{Vazan} A.,  {Sari} R.,   {Kessel} R.,  2022, \mn@doi [\apj] {10.3847/1538-4357/ac458c}, \href {https://ui.adsabs.harvard.edu/abs/2022ApJ...926..150V} {926, 150}

\bibitem[\protect\citeauthoryear{{Venturini}, {Guilera}, {Haldemann}, {Ronco}  \& {Mordasini}}{{Venturini} et~al.}{2020}]{Venturini2020}
{Venturini} J.,  {Guilera} O.~M.,  {Haldemann} J.,  {Ronco} M.~P.,   {Mordasini} C.,  2020, \mn@doi [\aap] {10.1051/0004-6361/202039141}, \href {https://ui.adsabs.harvard.edu/abs/2020A&A...643L...1V} {643, L1}

\bibitem[\protect\citeauthoryear{Vinet et~al.}{Vinet et~al.}{1989}]{Vinet1989}
Vinet P.,  et~al., 1989, J. Phys. Condensed Matter, 1, 1941

\bibitem[\protect\citeauthoryear{{Wagner} \& {Pru{\ss}}}{{Wagner} \& {Pru{\ss}}}{2002}]{Wagner2002}
{Wagner} W.,  {Pru{\ss}} A.,  2002, \mn@doi [Journal of Physical and Chemical Reference Data] {10.1063/1.1461829}, \href {https://ui.adsabs.harvard.edu/abs/2002JPCRD..31..387W} {31, 387}

\bibitem[\protect\citeauthoryear{{Wunderlich} et~al.,}{{Wunderlich} et~al.}{2019}]{Wunderlich2019}
{Wunderlich} F.,  et~al., 2019, \mn@doi [\aap] {10.1051/0004-6361/201834504}, \href {https://ui.adsabs.harvard.edu/abs/2019A&A...624A..49W} {624, A49}

\bibitem[\protect\citeauthoryear{{Yang}, {Cowan}  \& {Abbot}}{{Yang} et~al.}{2013}]{Yang2013}
{Yang} J.,  {Cowan} N.~B.,   {Abbot} D.~S.,  2013, \mn@doi [\apjl] {10.1088/2041-8205/771/2/L45}, \href {https://ui.adsabs.harvard.edu/abs/2013ApJ...771L..45Y} {771, L45}

\bibitem[\protect\citeauthoryear{{Yu}, {Moses}, {Fortney}  \& {Zhang}}{{Yu} et~al.}{2021}]{Yu2021}
{Yu} X.,  {Moses} J.~I.,  {Fortney} J.~J.,   {Zhang} X.,  2021, \mn@doi [\apj] {10.3847/1538-4357/abfdc7}, \href {https://ui.adsabs.harvard.edu/abs/2021ApJ...914...38Y} {914, 38}

\bibitem[\protect\citeauthoryear{{Zeng} \& {Sasselov}}{{Zeng} \& {Sasselov}}{2013}]{Zeng2013}
{Zeng} L.,  {Sasselov} D.,  2013, \mn@doi [\pasp] {10.1086/669163}, \href {https://ui.adsabs.harvard.edu/abs/2013PASP..125..227Z} {125, 227}

\bibitem[\protect\citeauthoryear{{Zeng} et~al.,}{{Zeng} et~al.}{2019}]{Zeng2019}
{Zeng} L.,  et~al., 2019, \mn@doi [Proceedings of the National Academy of Science] {10.1073/pnas.1812905116}, \href {https://ui.adsabs.harvard.edu/abs/2019PNAS..116.9723Z} {116, 9723}

\bibitem[\protect\citeauthoryear{{de Pater} \& {Lissauer}}{{de Pater} \& {Lissauer}}{2010}]{DePater2010}
{de Pater} I.,  {Lissauer} J.~J.,  2010, {Planetary Sciences}.
{Cambridge University Press}

\makeatother
\end{thebibliography}







\bsp	
\label{lastpage}
\end{document}